\documentclass[journal]{IEEEtran}
\usepackage{cite}
\usepackage[backref]{hyperref}
\usepackage{amsmath}

\usepackage{array}
\newcolumntype{C}[1]{>{\centering}p{#1}}
\ifCLASSINFOpdf
   \usepackage[pdftex]{graphicx}
   \graphicspath{{../pdf/}{../jpeg/}}
   \DeclareGraphicsExtensions{.pdf,.jpeg,.png}
\else
   \usepackage[dvips]{graphicx}
   \graphicspath{{../eps/}}
   \DeclareGraphicsExtensions{.eps}
\fi

\usepackage{caption}

\usepackage{multirow}
\usepackage{float}

\usepackage{graphicx}

\ifCLASSOPTIONcompsoc
  \usepackage[caption=false,font=normalsize,labelfont=sf,textfont=sf]{subfig}
\else
  \usepackage[caption=false,font=footnotesize]{subfig}
\fi

\begin{document}
\title{A New Radar Signal Multiparameter-Based Deinterleaving Method}

\author{Wang~Chao,
        Liu Weisong,
        Li Xueqiong,
        Wang Xiang,
        and~Huang~Zhitao

\thanks{Wang Chao is with College of Electronic Science and Technology and College of Electronic Engineering, National University of
Defense Technology, Changsha 410003, China, E-mail: (wangchaoben@126.com).}
\thanks{Liu Weisong, Wang Xiang, and Huang Zhitao are with College of Electronic Science and Technology, National University of 
Defense Technology, Changsha 410003, China, E-mail: (liuweisong15@nudt.edu.cn; christopherwx@163.com; huangzhitao@nudt.edu.cn).}
\thanks{Li Xueqiong is with College of Computer Science and Technology, National University of
Defense Technology, Changsha 410003, China, E-mail: (lixueqiong13@nudt.edu.cn).}

\thanks{Corresponding author: Wang Xiang.}
}


\maketitle
\begin{abstract}
Radar signal deinterleaving has been extensively and thoroughly investigated in the electronic reconnaissance field. In this work, a new radar signal multiparameter-based deinterleaving method is proposed. In this method, semantic information composed of the pulse repetition interval (PRI), pulse width (PW), radio frequency (RF),  and pulse amplitude (PA) of a radar signal is used to deinterleave radar signals. A bidirectional gated recurrent unit (BGRU) is employed, and the difference of time of arrival (DTOA)/RF, DTOA/PW, and DTOA/PA of the pulse stream are input into the BGRU. Based on the semantic information contained in different radar signal types, each pulse in the obtained pulse stream is classified according to the semantic information category, and the radar signals are deinterleaved. Compared to the PRI-based deinterleaving methods, the proposed method utilizes the multidimensional information of radar signals. As a result, higher deinterleaving accuracy is achieved. Compared to other existing radar signal multiparameter-based deinterleaving methods, the proposed method can adapt to radar signals with complex parameter features as well as to complex signal environments, and can complete the use of multiparameter in one step.

\end{abstract}


\IEEEpeerreviewmaketitle

\section{Introduction}
\begin{figure*}[tb]
\centering
\includegraphics[width=5.5in]{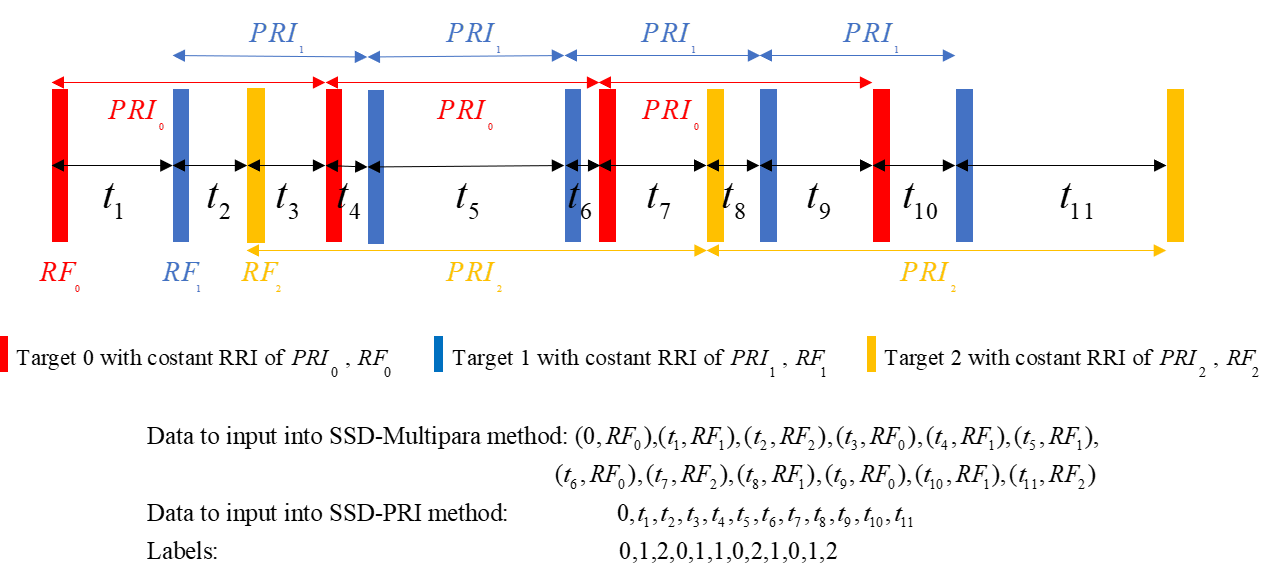}
\caption{Input and label of an interleaved pulse stream.}
\label{interleaved}
\end{figure*}

In modern warfare, intercepting electromagnetic radiation signals using electronic reconnaissance equipment is necessary to obtain the relevant information about the target radars. In an actual electromagnetic environment, multiple electromagnetic radiation sources often exist in the same area. Therefore, the data collected by the electronic reconnaissance equipment will contain information from different targets. In this case, the pulses obtained from different radiation sources constitute an interleaved pulse stream \cite{1993-Wiley}, which is intercepted by the reconnaissance equipment, as shown in Fig. \ref{interleaved}.

A pulse description word represents pulse description data, including the time of arrival (TOA), the direction of arrival (DOA), pulse width (PW), radio frequency (RF), pulse amplitude (PA), and other information. In electronic reconnaissance equipment, full pulse data (FPD) are pulse description words of each pulse output in chronological order. Radar signal deinterleaving, which is an important part of electronic reconnaissance, refers to deinterleaving the interleaved pulse description words in the FPD that belong to different radiation sources.

The research that has been conducted on radar signal deinterleaving can be divided into two categories: TOA-based deinterleaving and multiparameter-based deinterleaving. The former uses only the TOA of a pulse stream, whereas the latter comprehensively uses the TOA, DOA, PW, RF, PA, and other information.

The pulse repetition interval (PRI) is the interval between the fronts of adjacent pulses when a radar system transmits signals, as shown in Fig \ref{interleaved}. The TOA-based deinterleaving methods are also called the PRI-based deinterleaving methods. This kind of methods have been widely applied to deinterleaving radar signals with periodical PRIs. These methods exhibits good and stable performance, especially in deinterleaving radar signals with a constant PRI.

\begin{figure}[tb]
\centering
\includegraphics[width=2.5in]{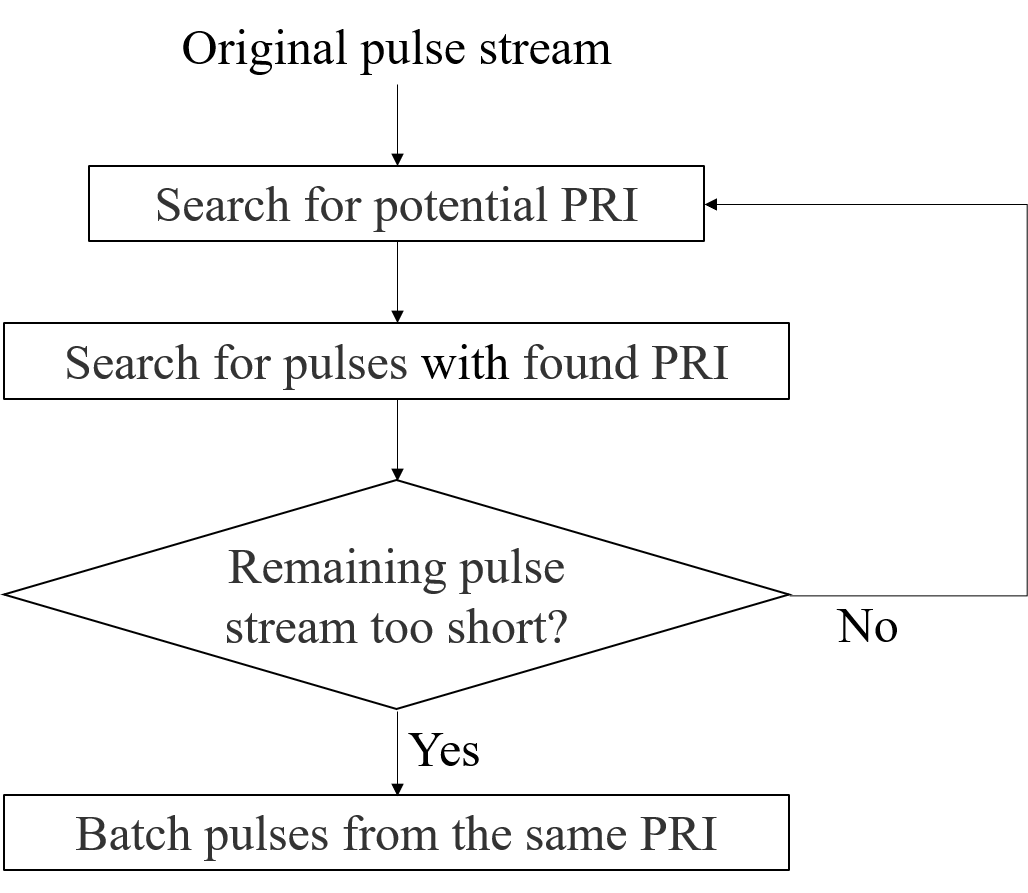}
\caption{Deinterleaving process of the traditional methods based on determining the PRI.}
\label{PRI_base_method}
\end{figure}

In the above methods, the most widely used concept is to first determine the radar PRI or PRI period from the pulse stream and then use the found PRI or PRI period to search for the target radar pulses from the pulse stream, as shown in Fig. \ref{PRI_base_method}. Some methods utilize the difference of TOA (DTOA) histogram, such as the cumulant difference histogram \cite{1989-CDIF} and the sequential difference histogram (SDIF) \cite{1992-SDIF} , to obtain the radar PRI and PRI period. Compared to the former, the latter methods significantly reduce the amount of computation and improve the practicability of the algorithm. Some other methods utilize the DTOA matrix to determine the PRI and PRI period \cite{2002-New-Matrix-Method} \cite{1998-A-Novel-Pulse-TOA-Analysis-Technique}. Another technique is to obtain the PRI spectrum through the transformation of TOA and extract the real PRI and PRI period by searching for the spectrum peak \cite{1999-Spectrum-Estimation-of-Interleaved-Pulse-Trains} \cite{2000-PRITran}. The PRI transform (PRI-Tran) algorithm has attracted significant attention due to its excellent harmonic suppression performance \cite{2000-PRITran}. However, the above-mentioned methods exhibit several disadvantages. First, when the radar signal pulses are dense or the target pulse loss rate is high, it is difficult to determine the right PRI and PRI period. Second, when searching for potential PRI and PRI period and target pulses, it is necessary to set thresholds and tolerances based on experience, making the deinterleaving efficiency prone to large fluctuations. In addition, regarding the threshold setting when searching for the PRI and PRI period, the existing algorithms only provide the theory for radar signals with a constant PRI; they do not provide the theory for radar signals with complex PRI modulation modes. Third, these methods require iterating the input and output data repeatedly to find a new PRI or PRI period and search pulses, which increases the complexity of the algorithm and requires additional computation time.

In other methods, a radar pulse stream with a periodical PRI is modeled as a linear dynamic model, and the Kalman filter is used for deinterleaving \cite{1994-Kalman} \cite{2010-Kalman-MHT} \cite{1999-Kalman} \cite{1998-Kalman}. Some research efforts have also been made to apply the hidden Markov model to radar signal deinterleaving \cite{2005-HiddenMarkov} \cite{2002-An-interval-amplitude-algorithm}. The algorithmic processes in these methods work only when some pre-assumptions are met. They are also complicated for practical applications.

\begin{figure}[!t]
\centering
\includegraphics[width=3in]{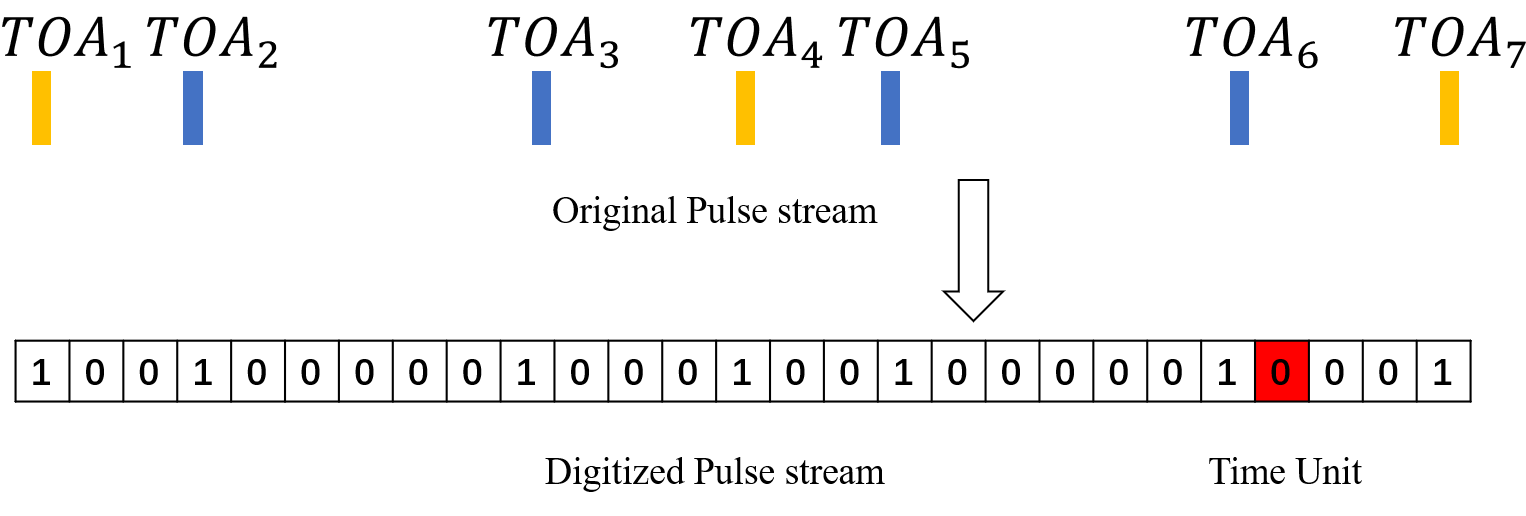}
\caption{TOA digitization.}
\label{TOA_digitization}
\end{figure}

\begin{figure}[hbpt]
\centering
\includegraphics[width=2.5in]{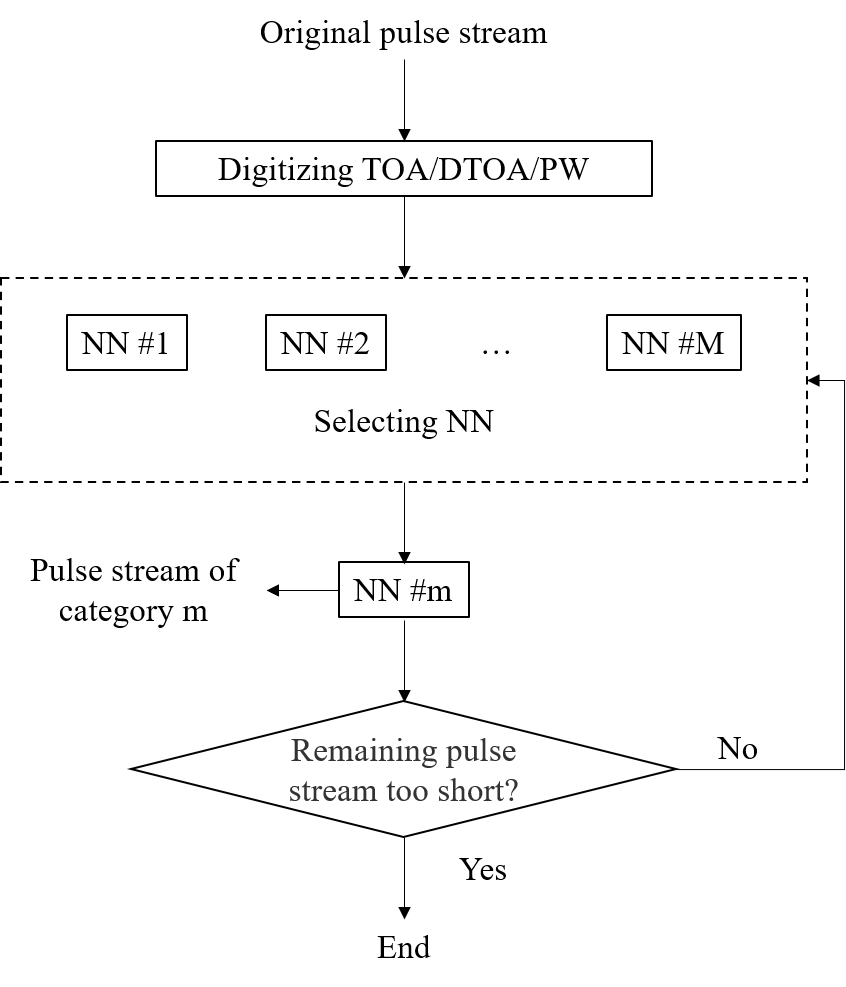}
\caption{Deinterleaving process performed by existing methods using NNs.}
\label{NN_method}
\end{figure}

Early efforts employing neural networks (NNs) to classify pulses have also been reported \cite{2013-Radar-Emitter-Signals-Recognition-and-Classification-with-Feedforward-Networks}  \cite{1998-self-organizing-networks-clustering-pulses} \cite{1991-Radar-signal-categorization-using-a-neural-network} \cite{2007-Deinterleaving-of-radar-signals-and-PRF-identification-algorithms}. Recurrent NNs (RNNs) were introduced to deinterleave radar signals in \cite{2020-lzm-Classification-Denoising}. In that study, the deinterleaving problem was treated as a prediction problem. Thus, only the unidirectional information related to a pulse sequence is used to assess the category of each pulse. Autoencoders have been used to reduce noise pulses, but they are not capable of finding the target pulses \cite{2020-Denoising-Autoencoders}. In this case, accurate prior information about the target pulse parameters is required \cite{2020-Deinterleaving-Autoencoders}. To facilitate NN processing of radar FPD, the methods reported in \cite{2020-lzm-Classification-Denoising} \cite{2020-Denoising-Autoencoders} \cite{2020-Deinterleaving-Autoencoders} use small time units to digitize time information, such as TOA, DTOA, PRI, and PW, as shown in Fig. \ref{TOA_digitization}. This operation introduces three problems, which we call “resolution problems.” First, it introduces errors and reduces the time information accuracy. Second, when there is more than one pulse in a time unit, only one pulse is presented, and information about other pulses is covered. Third, the methods based on this operation may classify a no-pulse position as a pulse position. In addition, the above methods need to train an NN for each radar signal category, and only one target can be deinterleaved in each output step, as shown in Fig. \ref{NN_method}. In other words, the existing methods complete a binary classification task in each output step. Therefore, it is necessary to iterate the input and output data repeatedly. Finite automata have also been used for radar signal deinterleaving, but they also require prior information about the target pulse parameters \cite{2020-Deinterleaving-Automata}.

\begin{figure}[!t]
\centering
\includegraphics[width=3in]{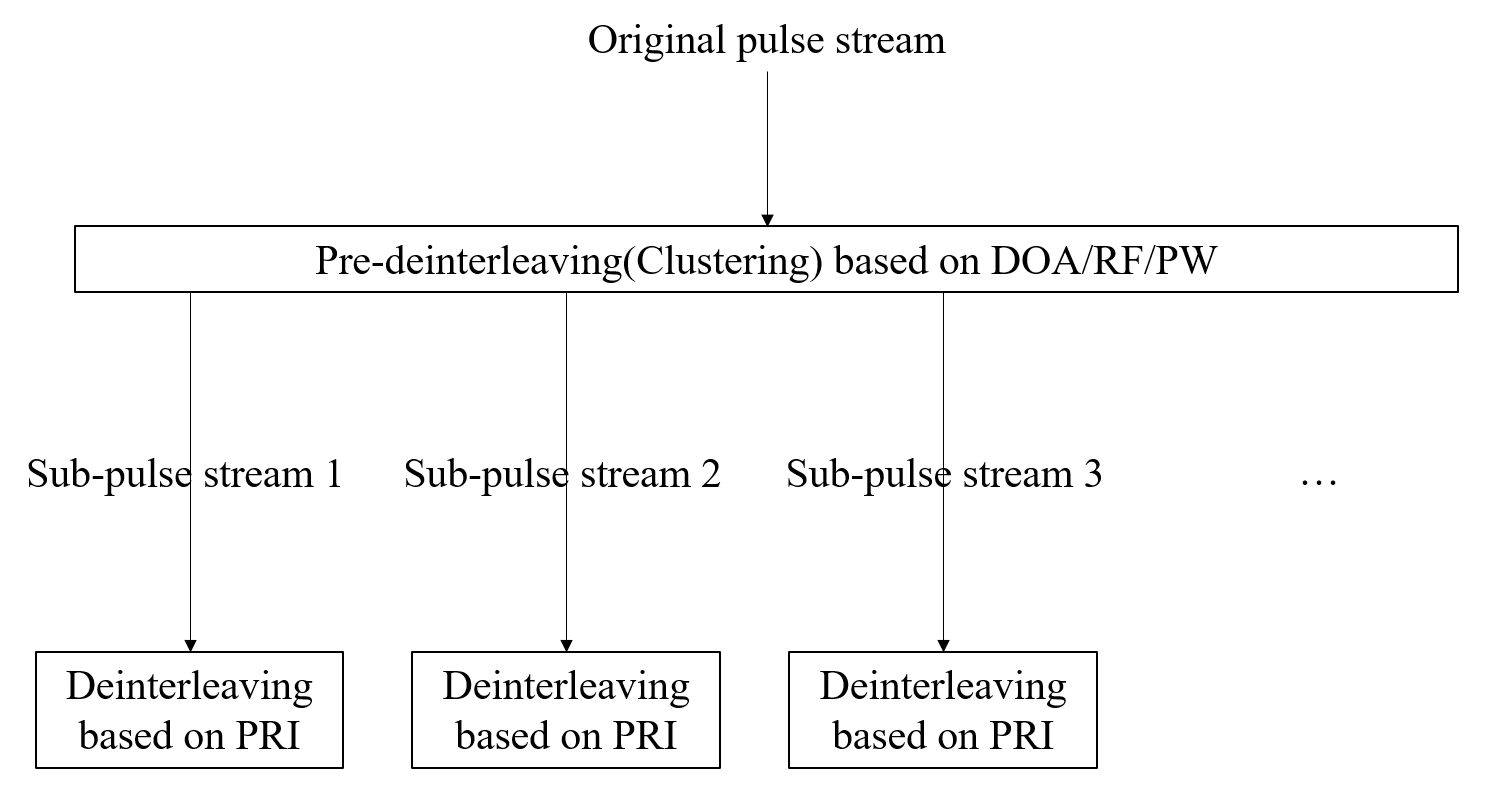}
\caption{Deinterleaving process used in existing multiparameter-based deinterleaving methods.}
\label{multi_base_method}
\end{figure}

In radar signal multiparameter-based deinterleaving, the DOA, PW, and RF are typically used to cluster the radar pulses first. This is also known as pre-deinterleaving. Then, the deinterleaving methods based on PRI are used to further process the clustered results to further improve the deinterleaving efficiency. The overall deinterleaving process is presented in Fig. \ref{multi_base_method}.

For multiparameter-based deinterleaving methods, different parameters have been used and different clustering methods have been proposed  in the existing literatures. The DOA-based approach was used in \cite{1982-deint-multipara-doa} \cite{2003_Joint_deinterleaving/recognition_of_radar_pulses}, the RF/DOA-based approach was used in \cite{1985-deint-multipara-rf}, the RF/PW-based approach was used in \cite{2020_Clustering_radar_pulses_with_Bayesian_nonparametrics} \cite{2021_Deinterleaving_and_Clustering_unknown}, and the PW/RF/DOA-based approach was used in \cite{2019-deint-multipara-hit} \cite{2016_On_cluster_validity_indices_with}. 

These multiparameter-based methods are expected to provide better results because they use more information. However, they exhibit several disadvantages. First, \textbf{they are not applicable to radar signals with variable parameters}. The function of modern radar systems is diverse, and their technology is complex. The operating parameters of a radar system often vary significantly in a short time. Thus, pulses belonging to the same target are often clustered into multiple targets. For this reason, some researchers only use the DOA for clustering since the radar position cannot change in a short time \cite{1982-deint-multipara-doa} \cite{2003_Joint_deinterleaving/recognition_of_radar_pulses}. The support vector machine, which is based on the PRI/RF/PA approach, is used to deinterleave radar signals with variable parameters, but it is limited to specific scenarios \cite{2019-deint-multipara}. Second, \textbf{these methods are not applicable to cases where the parameter values of multiple target radars overlap}. In a real environment, the reconnaissance equipment often receives signals from multiple targets simultaneously, which are in the same direction and operating in the same parameter value range. The clustering-based deinterleaving method considers different targets with the same parameter value range as the same one. \cite{2019-deint-multipara-hit}. Third, \textbf{these methods sever the relationship between the parameters}. As shown in Fig. \ref{multi_base_method}, the deinterleaving process is divided into two steps: pre-deinterleaving based on RF/DOA/PW and deinterleaving based on PRI. The deinterleaving process needs to be completed step by step. However, some radar signal features are composed of different parameters, such as duty cycle (DC) obtained by PW/PRI. The stepwise deinterleaving methods break the relationship between different parameters and does not make full use of the hidden features of the pulse stream.

\begin{figure}[!t]
\centering
\includegraphics[width=2.5in]{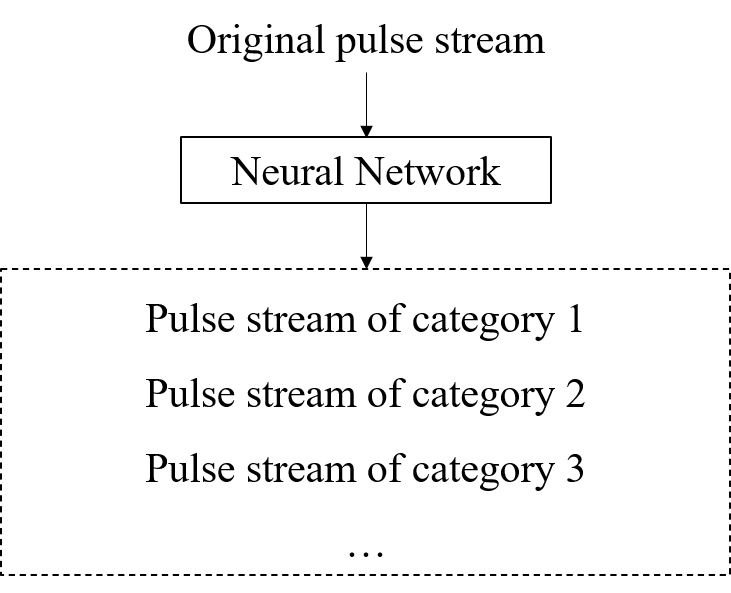}
\caption{Deinterleaving process of the SSD method.}
\label{SSD_method}
\end{figure}

In our previous study \cite{ssd-pri}, we proposed a new radar signal deinterleaving method referring to semantic segmentation. We called this method the “semantic segmentation deinterleaving (SSD) method.” We take the PRI characteristics of radar signals as semantics. Each pulse in the interleaved pulse stream is labeled by an NN according to the category of semantic contained. In this way, the deinterleaving process of different radar signal types can be realized. This method is capable of deinterleaving the signals of multiple target radars using one NN in one step, as shown in Fig. \ref{SSD_method}. Compared to traditional PRI-based methods and existing deinterleaving methods employing NNs, the proposed SSD method overcomes their disadvantages, and excellent deinterleaving results can be obtained. Although much progress has been made on the SSD method, there are still some limitations. First, \textbf{when multiple target radars have the same PRI modulation modes and the PRI value ranges overlap}, this method cannot classify the target radar signals. Second, \textbf{this method only uses the TOA of signals}. To further improve the deinterleaving accuracy, more parameters are required.

In this paper, we continue to use the concept of semantic segmentation to study the radar signal multiparameter-based deinterleaving method, which we call “SSD multiparameter-based (SSD-Multipara) method.” We call the previous method \cite{ssd-pri} “SSD PRI-based (SSD-PRI) method.” The new method is used here to overcome the limitations of the SSD-PRI method and the disadvantages of existing multiparameter-based deinterleaving methods.

Compared to the PRI-based methods, \textbf{this method uses more parameters, and can get much more excellent deinterleaving efficiency}.

Compared to the SSD-PRI method, \textbf{this method can adapt to the situation where the PRI features of multiple radars are the same}.

Compared to existing multiparameter-based deinterleaving methods, \textbf{this method 
is suitable for radar signals with variable parameters and cases where the parameter values of different targets overlap, and can use all parameters in one step}.

This paper is organized as follows. The task characteristics and the proposed method are introduced in Section II. In Section III, the NN is selected, and the data model is defined. The simulation experiments and the results are presented in Section IV. Section V concludes the entire paper.

\section{TASK CHARACTERISTICS AND THE PROPOSED METHOD}
\subsection{Characteristics of radar signal parameters}
Each parameter of the target radars contains the corresponding information. When this information can be effectively used for deinterleaving, the deinterleaving accuracy can be further improved. However, the complex characteristics of modern radars and their operating environments make it difficult to use these parameters. 
Since deinterleaving is based on the FPD intercepted by the reconnaissance equipment, the characteristics of the radar operating parameters are discussed based on the data intercepted by the reconnaissance equipment in this paper. 
These characteristics are described below.

\subsubsection{Complex variation mode  and wide variation range}
To implement a diverse function, the short-time variation of the operating parameters of a modern radar system must be considered. A radar system with frequency agility operates in a frequency-hopping state in the RF band, which enhances its anti-jamming ability. The PW variation allows a radar system to achieve a balance between detection precision and range. The variation of the radar detection direction results in the variation of the radar signal amplitude in the reconnaissance equipment. The signal amplitude of a mechanical-scanning radar system varies continuously, while that of a phased array radar system varies step by step.

\subsubsection{Overlapping of parameter values of different target}
In a real environment, there are usually multiple radars and other electromagnetic radiation sources in the same area. This makes it possible to have multiple targets in the same direction in the reconnaissance equipment, and difficult to distinguish the different targets from the direction. Modern radars typically have a large range of operating parameter values. When signal parameter values from different targets intercepted by the reconnaissance equipment overlap, these signals become indistinguishable.

\subsubsection{Multiparameter-composed target characteristics}
Some radar signal characteristics are composed of different parameters. This kind of characteristics reflects the relationship between different parameters and cannot be extracted form a single parameter. For example, two different radars may have the same PRI modulation mode, PRI value range, and PW variation mode, and their PW value range overlap, but they adopt different DCs, which is given by dividing PRI by PW. The scanning mode of radar is reflected by TOA and PA, which cannot be obtained by PA or TOA alone. In many cases, the different parameters are coupled.

\subsection{Advantages of NNs in multiparameter utilization}
The advantages of NNs in multiparameter utilization are described below.
\subsubsection{Ability of NNs to extract complex features from radar parameters due to their strong fitting ability}
Data processing methods based on clustering are capable of classifying the data using the distance between samples. However, these methods cannot extract complex data features. When the parameter values of a single target vary in a wide range or multiple targets operate in the same parameter value range, clustering errors occur. However, NNs are capable of extracting the principle governing the data variation process due to their strong fitting ability. A good fitting can be obtained even for complex signal parameters.

\subsubsection{Ability of NNs to extract features from multiple parameters by allowing the simultaneous input of multiple parameters}
The existing radar signal multiparameter-based deinterleaving methods usually deal with different FPD parameters on a step-by-step basis. This approach disconnects the relationship between different radar signal parameters. In particular, the TOA must be used after pre-deinterleaving, which obscures the time-varying features of other parameters. NNs allow the simultaneous input of multiple parameters, making the extraction of features contained in multiple parameters easy.

\subsection{Input and output data form}
In the SSD-PRI method, $1*N$ dimension data are input into an NN, whereas, in the SSD-Multipara method, $x*N$  dimension data are input into an NN. $x$ represents the multidimensional description information of each pulse in the pulse stream, where one dimension is the TOA and the other dimensions can be PW, RF, or PA. In this paper, $x$ is 2. $N$ represents the length of the pulse stream, which is the number of pulses in the pulse stream. . 

In the proposed method, the DTOA of the pulse stream is input into the NN instead of the TOA. In this way, the input data exhibit smaller variance and facilitate the NN processing. Thus, the inputs are DTOA/RF, DTOA/PW, and DTOA/PA. Fig. 1 shows the case where the input is DTOA/RF.

When the pulse stream contains only a single radar target, the DTOA of the pulse stream is the real PRI of the target. When the pulse stream contains pulses from multiple radiation sources or pulses from a single target arrive through multipath propagation, the DTOA of the pulse stream is chaotic. To make DTOA and TOA equal in length, 0 is added before DTOA as the first value of DTOA.

Different from DTOA, the PW, RF, and PA only describe the features of a single pulse and have no relationship with the adjacent pulses. When a target radar pulse is lost, or the pulses from different radiation sources are interleaved, the PW, RF, and PA of the intercepted pulses will not change, but the DTOA input at these pulse points will change.

In the proposed method, the output represents the label information of each pulse. During the NN training, the description word and label of each pulse are input into the NN, as shown in Fig. \ref{interleaved}.

\subsection{Differences between the proposed tasks and image semantic segmentation and other sequence modeling tasks}
Radar signal deinterleaving based on semantic segmentation is a problem of mapping an input sequence to an output sequence. It is different from image semantic segmentation \cite{2017-DeepLabv2} \cite{2017-SegNet} and sequence-to-sequence (seq2seq) tasks such as natural language processing \cite{2014-Neural-Machine-Translation-Jointly-Learning-Align-and-Translate} \cite{2016-Neural-Machine-Translation-in-Linear-Time}. These differences are described below.
\subsubsection{Unconcentrated target points that run throughout the sequence}
In the pulse stream, the pulses of different targets are interleaved and information about the same target runs through the entire pulse stream. However, in image semantic segmentation, the pixels of the same object are typically concentrated in one or several regions.

\subsubsection{Strict mathematical relationship between the data at each input point of the sequence}
Since the input contains the DTOA, the information loss of one data point completely changes the information in the pulse stream. Therefore, pooling is not allowed. Image semantic segmentation and seq2seq tasks do not have this feature.

\subsubsection{Input and output sequences of equal length}
The input and output sequences of this task have equal lengths. In some sequence modeling tasks, such as machine translation, the input and output are often not of equal length.

\subsubsection{Different meanings of the multiple dimensions of data for each point; the multiple dimensions constitute new meanings}
The DTOA input data alone at each point is meaningless. Only when they are computed together with the input DTOA data before and after  them can their information be reflected. PW, RF, and PA all have their own meanings. The combination of the two dimensions of input data with different meanings describes how the PW, RF, and PA of pulses vary with time and constitutes complete information about a pulse stream.

\subsubsection{Equivalent forward and backward information}
In the deinterleaving task, the forward and reverse information of a sequence are completely equivalent, which is significantly different from many sequence modeling tasks. Therefore, in this task, it is more conducive to accurately estimate the category of each pulse using the bidirectional information of the sequence simultaneously.

\section{NN AND DATA MODELS}	
\subsection{Appropriate selection of NN type for the proposed task}
\begin{figure}[!t]
\centering
\includegraphics[width=3in]{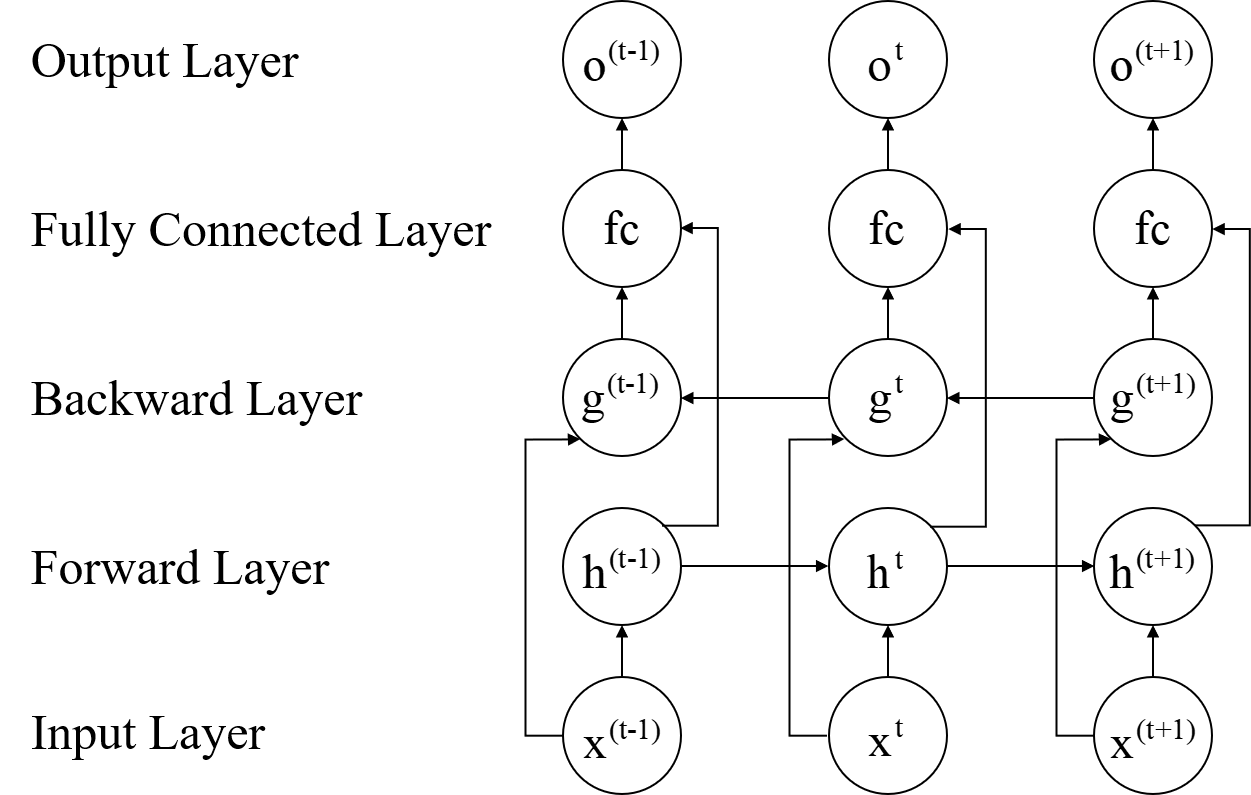}
\caption{Deinterleaving process of BGRU.}
\label{BGRU}
\end{figure}

In previous studies on the SSD-PRI method, we have selected bidirectional RNNs \cite{1997-BRNN} \cite{1997-LSTM} \cite{2014-GRU} and dilated convolutional NNs \cite{2018-TCN-RNN} \cite{2016-WaveNet}, which exhibit certain advantages in handling sequence modeling tasks and have confirmed the advantages of the classical RNN architecture. Continuing our previous studies, in this paper, we use a bidirectional gated recurrent unit (BGRU) \cite{2014-GRU} to achieve the deinterleaving task. The output of each step of the BGRU is connected to the fully connected layer to implement the classification of the output of each time step, as shown in Fig. \ref{BGRU}.
The parameters of BGRU are presented in Table \ref{Para BGRU}.

\begin{table}[]
\centering
\caption{Parameters of BGRU}
\label{Para BGRU}
\begin{tabular}{|c|c|c|c|}
\hline
Input size & Hidden layer size & \begin{tabular}[c]{@{}c@{}}Number of\\ hidden layers\end{tabular} & Outputsize \\ \hline
2          & 120               & 3                                                                 & 4          \\ \hline
\end{tabular}
\end{table}

\subsection{Loss function}
In this task, the prediction loss by the NN for each sample is the average of the prediction loss of all pulses in the pulse stream, i.e.,
\begin{equation}
Loss = \frac{1}{N}\sum\nolimits_{n = 1}^N {los{s_n}}\label{Loss}.
\end{equation}
$los{s_n}$ is the prediction loss by the NN for the nth pulse. We use a cross-entropy loss function to evaluate the NN’s prediction performance for each pulse. This is described as follows:
\begin{equation}
loss =  - \sum\nolimits_{c = 1}^C {{P_c}\log (\mathop {{P_c}}\limits^ \wedge  )} \label{loss}.
\end{equation}
$C$ represents the category number of radar signals in the pulse stream; ${P_c}$ denotes whether the current pulse belongs to the cth radar signal category, and its value is either 0 or 1; $\mathop {{P_c}}\limits^ \wedge$ represents the probability that the current pulse belongs to the cth radar signal category in the NN’s prediction.

\subsection{Data model}

In this section, we model the PRI modulation modes, the RF variation modes, the PW variation modes, and the PA variation modes used in this paper.

\subsubsection{PRI}

In this paper, the following three PRI modulation modes are defined and simulated.

\paragraph{Constant PRI}

The radar PRI remains constant, and the PRI sequence can be represented as follows:
\begin{equation}
PR{I_n} = PR{I_0},n = 1,2,3...\label{Constant_PRI}.
\end{equation}
Constant PRI is shown in Figs. \ref{exp3_1} and \ref{exp4_1}.

\paragraph{Dwell and switch (D\&S) PRI}
Several PRIs form a group. The number of PRIs in each group is the same, and the PRI values vary periodically among groups. This can be mathematically expressed as follows:
\begin{equation}
PR{I_n} = PR{I_{n + j}},0 \le j < J\label{DSPRI_1},
\end{equation}
\begin{equation}
PR{I_n} = PR{I_{n + J*K}}\label{DSPRI_2}.
\end{equation}
$PR{I_n}$ is the first PRI in each group, $J$ represents the number of pulses in each group, and $K$ represents the number of pulse groups in one period, i.e., the number of PRI values in one period.
D\&S PRI is shown in Figs. \ref{exp1_1}, \ref{exp1_3}, \ref{exp1_5}, \ref{exp2_3}, \ref{exp2_5}, \ref{exp3_5}, and \ref{exp4_5}.
  
\paragraph{Staggered PRI}

The radar PRI consists of several fixed values and varies periodically. The PRI sequence can be described as follows:
\begin{equation}
PR{I_n} = PR{I_{n + L}},n = 1,2,3...\label{Staggered_PRI}.
\end{equation}
$L$ represents the number of PRI values in one period.
Staggered PRI is shown in Figs. \ref{exp3_3} and \ref{exp4_3}.

\subsubsection{RF}

The following three RF variation modes are defined and simulated.

\paragraph{Constant RF}

The radar RF remains constant. This can be mathematically expressed as follows:
\begin{equation}
R{F_n} = R{F_0},n = 1,2,3...\label{Constant_RF}.
\end{equation}
Constant RF is shown in Fig. \ref{exp3_4}.
	 
\paragraph{Agile RF among pulses}

The radar RF consists of a number of fixed values that vary periodically among pulses. This can be mathematically expressed as follows:
\begin{equation}
R{F_n} = R{F_{n + M}},n = 1,2,3...\label{RF_pulses_1}.
\end{equation}
$M$ represents the number of RF values in one period.
Agile RF among pulses is shown in Fig. \ref{exp3_2}.
  
\paragraph{Agile RF among pulse groups}

Several pulses with the same RF value form a pulse group. The number of pulses in each group is the same, and the RF values vary periodically among groups. This can be mathematically expressed as follows:
\begin{equation}
R{F_n} = R{F_{n + o}},0 \le o < O\label{RF_groups_1},
\end{equation}
\begin{equation}
R{F_n} = R{F_{n + O*P}}\label{RF_groups_2}.
\end{equation}
$R{F_n}$ is the RF value of the first pulse in each group, $O$ is the number of pulses in each group, and $P$ is the number of pulse groups in one period, i.e., the number of RF values in one period.
Agile RF among pulse groups is shown in Fig. \ref{exp3_6}.

\subsubsection{PW}

The following two PW variation modes are defined and simulated.
\paragraph{Constant PW}

The radar PW remains constant. This can be mathematically expressed as follows:
\begin{equation}
P{W_n} = P{W_0},n = 1,2,3...\label{Constant_PW}.
\end{equation}
Constant PW is shown in Fig. \ref{exp2_2}.
	 
\paragraph{D\&S PW}

Several pulses with the same PW value form a pulse group. The number of pulses in each group is the same, and the PW values vary periodically among groups. This PW variation occurs when the PRI employs the D\&S modulation mode. In this case, \textbf{the PW and PRI vary simultaneously and have the same period}. This can be mathematically expressed as follows:
\begin{equation}
P{W_n} = P{W_{n + q}},0 \le q < Q\label{DSPW_1},
\end{equation}
\begin{equation}
P{W_n} = P{W_{n + Q*R}}\label{DSPW_2}.
\end{equation}
$P{W_n}$ is the PW value of the first pulse in each group, $Q$ represents the number of pulses in each group, and $R$ represents the number of pulse groups in one period.
D\&S PW is shown in Figs. \ref{exp2_4} and \ref{exp2_6}.

\subsubsection{PA}
The PA variation is closely related to the radar scanning mode. In this paper, the PA variation is defined and simulated assuming three scenarios: radar non-scanning, radar mechanical-scanning and radar phase-scanning.

\paragraph{Radar non-scanning}
When the radar is not scanning, the PA is stable and can be described as follows:
\begin{equation}
P{A_n} = P{A_0},n = 1,2,3...\label{non-scan}.
\end{equation}
The PA of Radar non-scanning is shown in Fig. \ref{exp4_4}.
 
\paragraph{Radar mechanical-scanning}

When the radar scans mechanically, the PA exhibits an envelope, which fluctuates with the radar rotation. The envelope corresponds to the radiation pattern of the radar antenna. Due to the diversity of the radar antenna radiation patterns, the effect of the main lobe of the mechanical-scanning radar is only considered in this paper and the sinc function is used to describe the antenna main lobe of the mechanical-scanning radar. The PA can be described as follows:
\begin{equation}
P{A_n} = 10\log \frac{{\sin (a{\theta _n})}}{{a{\theta _n}}},n = 1,2,3...\label{mechanical-scan}.
\end{equation}
$a$ denotes the beamwidth coefficient, and ${\theta _n}$ denotes the angle between the direction of the incident wave from the target signal and the direction of the antenna main lobe.
The PA of Radar mechanical-scanning is shown in Fig. \ref{exp4_2}.
  
\paragraph{Radar phase-scanning}
When the radar operates in the phase-scanning mode, the PA varies instantly with the variation of the radar beam pointing, showing a step change. This can be described as follows:
\begin{equation}
P{A_n} = P{A_{n + r}},0 \le r < R\label{phase-scan},
\end{equation}
\begin{equation}
P{A_n} = P{A_{n + S*R}}\label{phase-scan}.
\end{equation}
$P{A_n}$ denotes the amplitude of the first pulse in each beam position, $R$ denotes the number of pulses in each beam position, and $S$ denotes the number of beam positions in one period. \textbf{In general, there are several groups of PRIs in each beam position}. In this paper, $T$ is used to represent the number of PRI groups in a beam position.
The PA of Radar phase-scanning is shown in Fig. \ref{exp4_6}.

\section{EXPERIMENTS}
\subsection{Data simulation}
Based on the above data model definition, the following are considered in the simulation:

1) In all experiments, the length of the pulse stream is 1000, i.e., the dimension of the data input into the BGRU is $2*{\rm{1000}}$.

2) The Gaussian distributed deviation is added to the TOA, RF, and PW to simulate the measurement errors, and the standard deviation is 0.1. The DTOA generation is based on the TOA. The Gaussian distributed deviation is added to the PA measurement voltage to simulate the measurement errors, and the standard deviation is 0.003. Based on this, the decibel values are generated and used in the experiments.

3) The problems of target pulse loss and random noise pulses in the intercepted pulse stream are considered. The pulse loss rate of the target is represented by ${\rho _l}$. Noise to target ratio, i.e., the ratio of the number of random noise pulses to the average number of the target radar pulses in the intercepted pulse stream is represented by ${\rho _n}$. Noise pulses rate, i.e., the proportion of the number of random noise pulses to the total number of pulses can be calculated by $\frac{{{\rho _n}}}{{{\rho _n} + D}}$, where $D$ represents the number of target radars.

\subsection{Design of experiments}
\begin{figure}[!t]
\centering
\subfloat[PRI features of target 1.]{\includegraphics[width=1.6in]{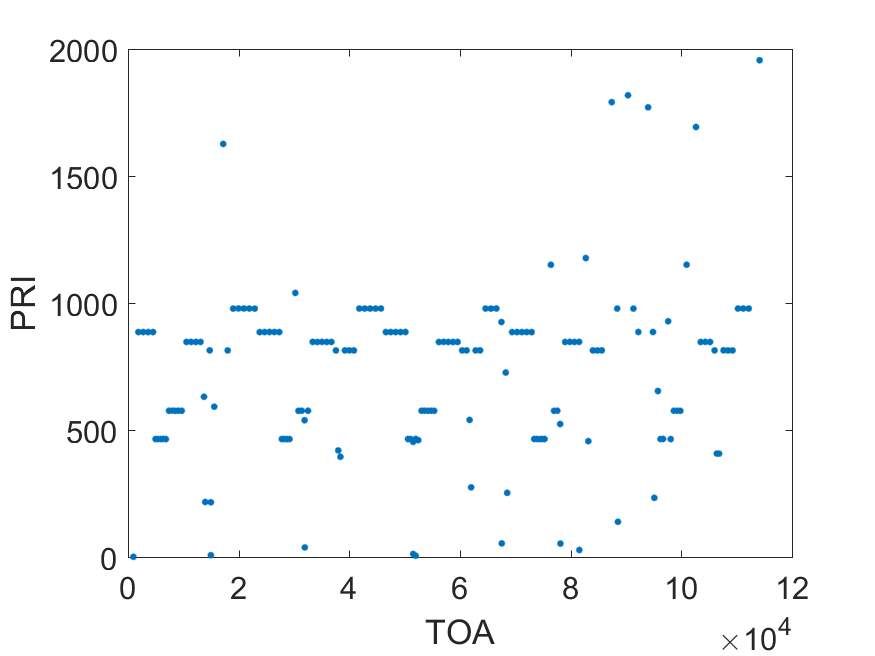}%
\label{exp1_1}}
\hfill
\subfloat[PW features of target 1.]{\includegraphics[width=1.6in]{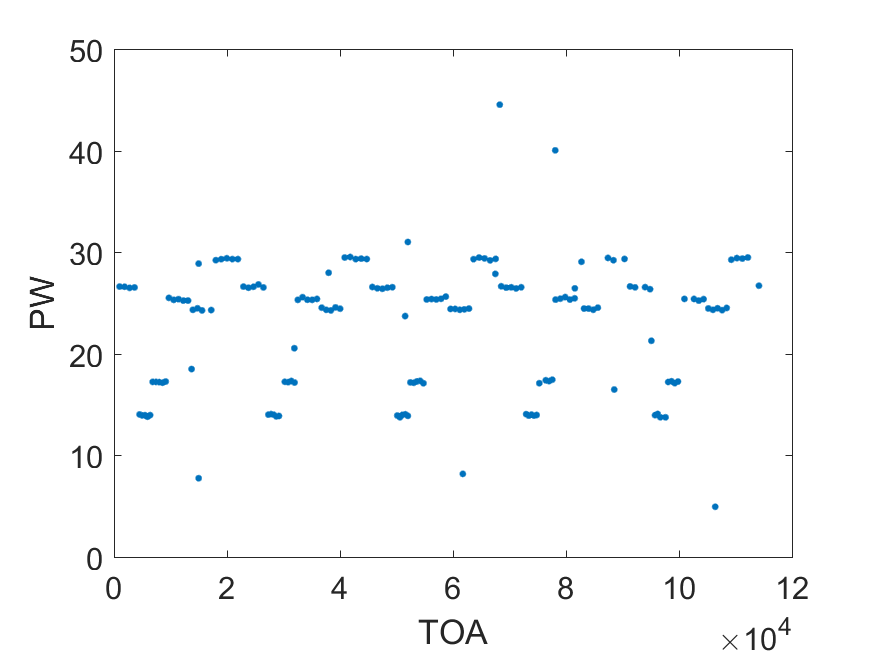}%
\label{exp1_2}}
\hfill
\subfloat[PRI features of target 2.]{\includegraphics[width=1.6in]{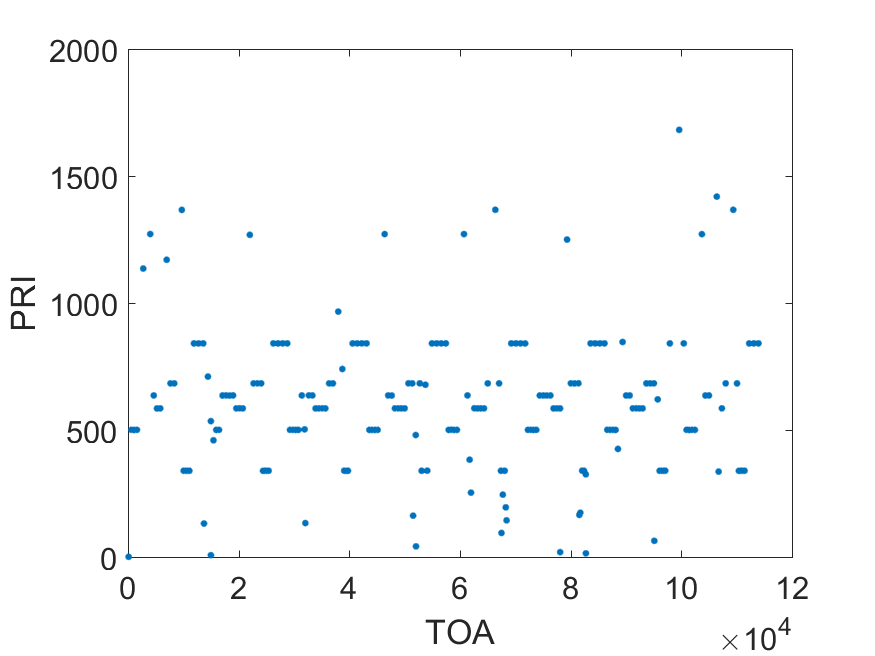}%
\label{exp1_3}}
\hfill
\subfloat[PW features of target 2.]{\includegraphics[width=1.6in]{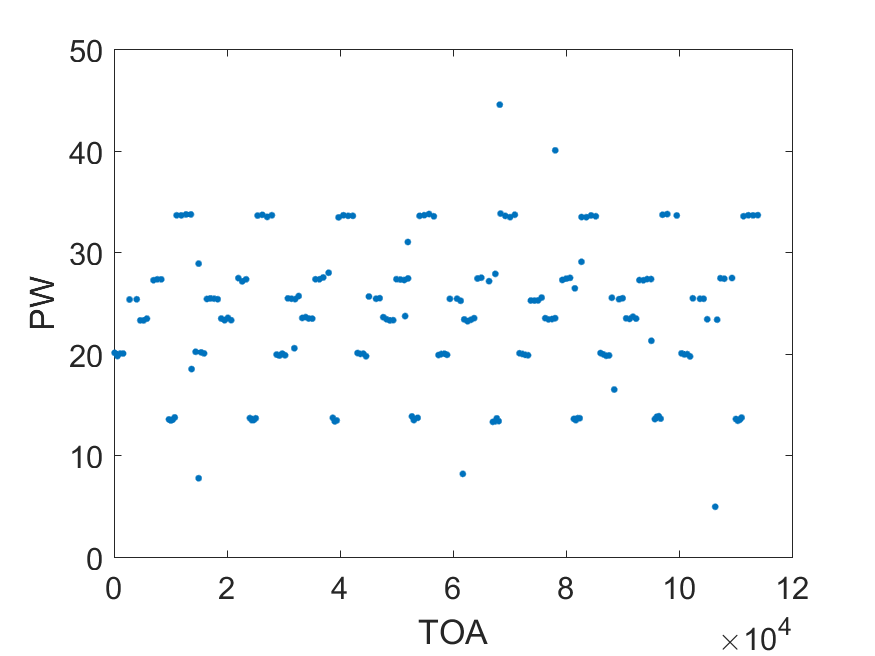}%
\label{exp1_4}}
\hfill
\subfloat[PRI features of target 3.]{\includegraphics[width=1.6in]{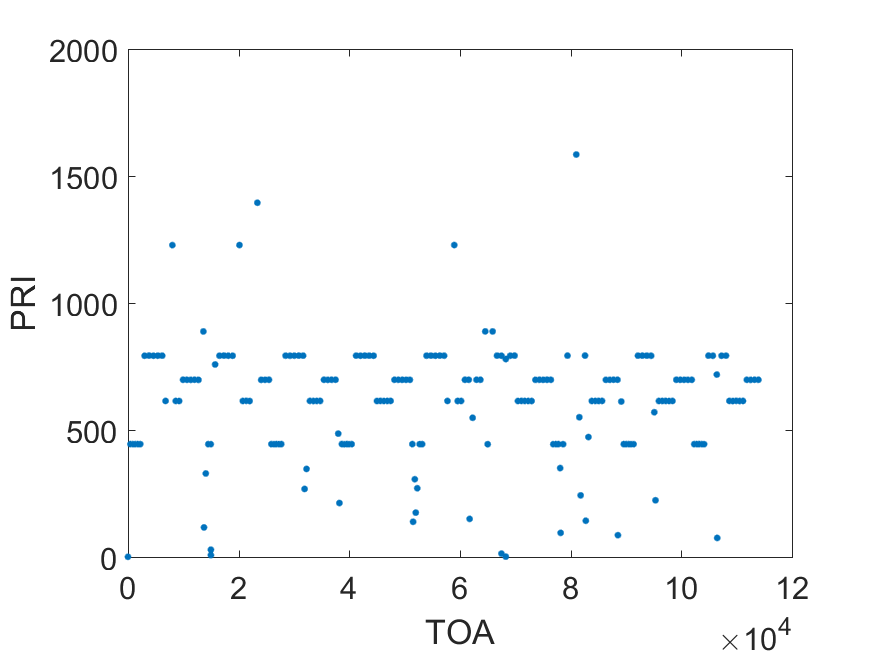}%
\label{exp1_5}}
\hfill
\subfloat[PW features of target 3.]{\includegraphics[width=1.6in]{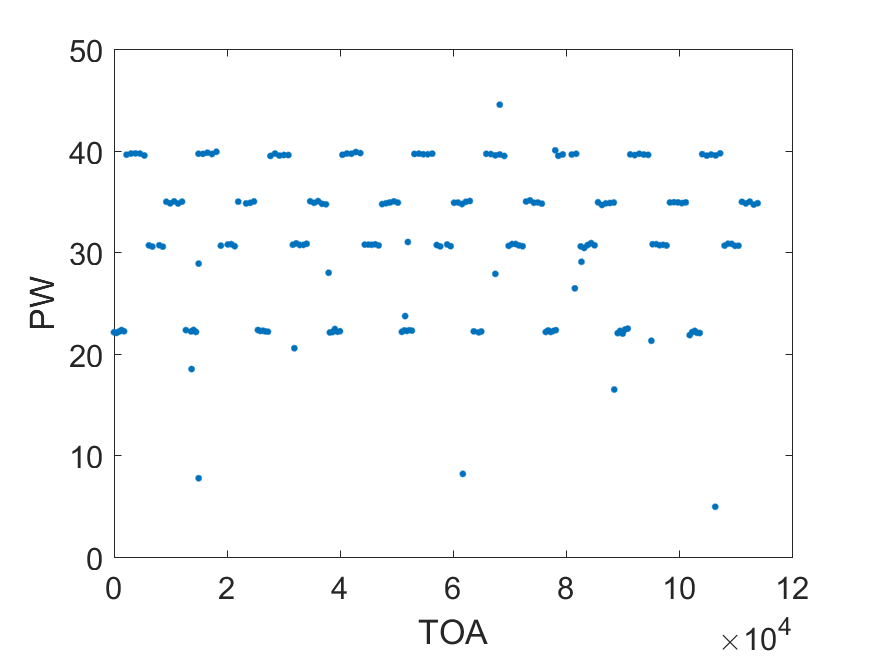}%
\label{exp1_6}}
\hfill
\subfloat[DTOA of interleaved pulse stream in Experiment 1.]{\includegraphics[width=1.6in]{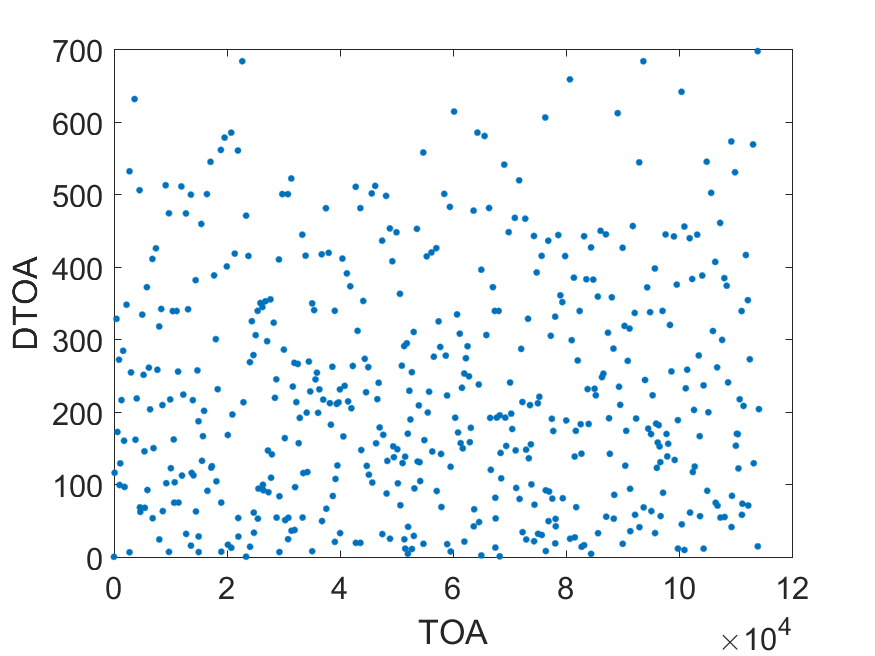}%
\label{exp1_7}}
\hfil
\subfloat[PW of interleaved pulse stream in Experiment 1.]{\includegraphics[width=1.6in]{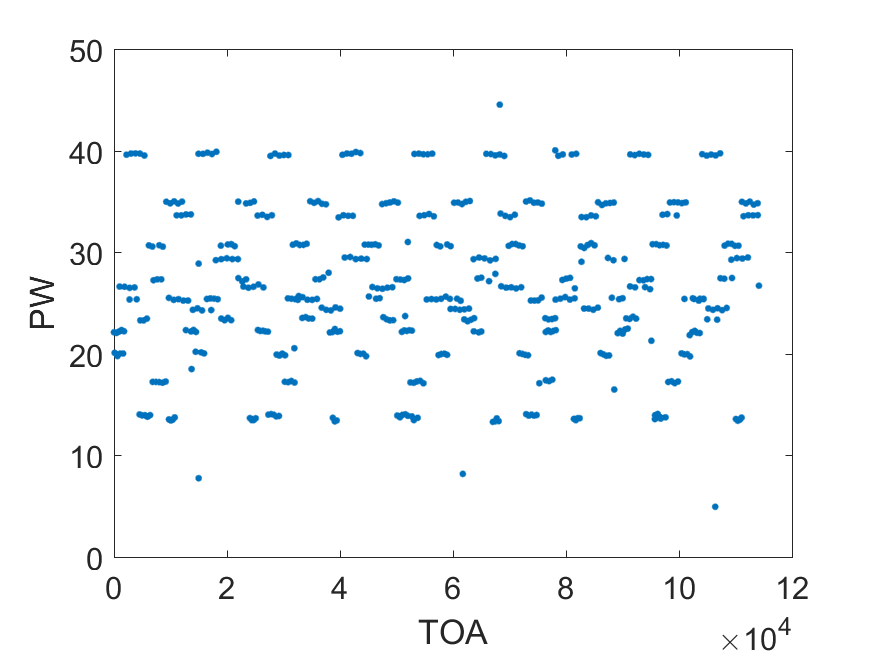}%
\label{exp1_8}}
\caption{Signal features of target radars in Experiment 1.}
\label{exp1}
\end{figure}

\begin{figure}[!t]
\centering
\subfloat[PRI features of target 1.]{\includegraphics[width=1.6in]{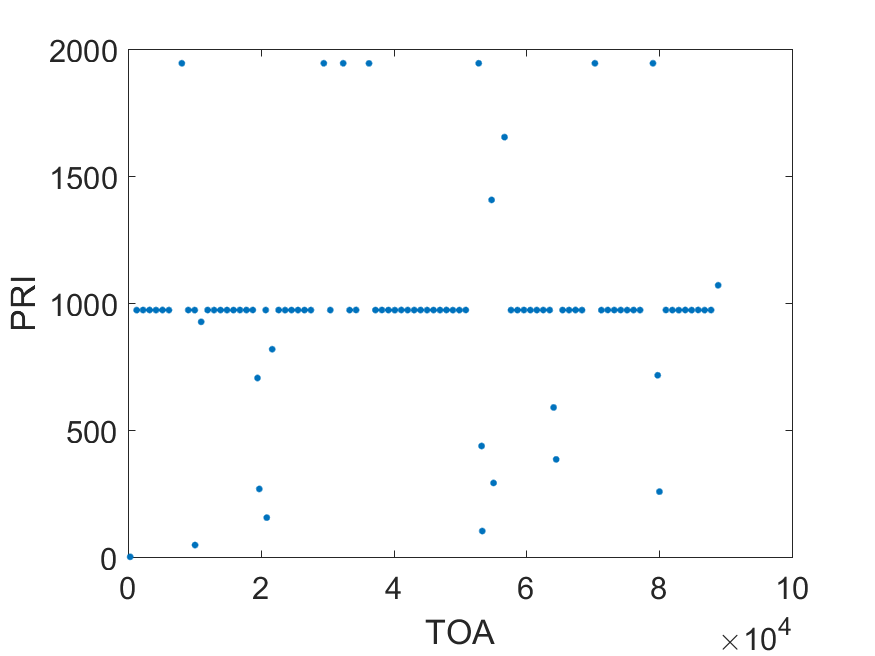}%
\label{exp2_1}}
\hfill
\subfloat[PW features of target 1.]{\includegraphics[width=1.6in]{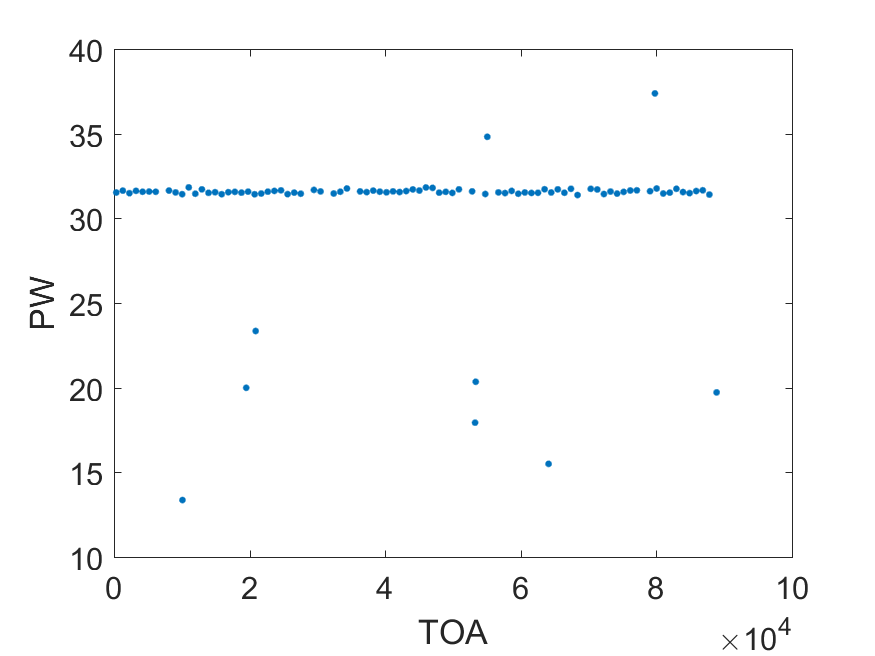}%
\label{exp2_2}}
\hfill
\subfloat[PRI features of target 2.]{\includegraphics[width=1.6in]{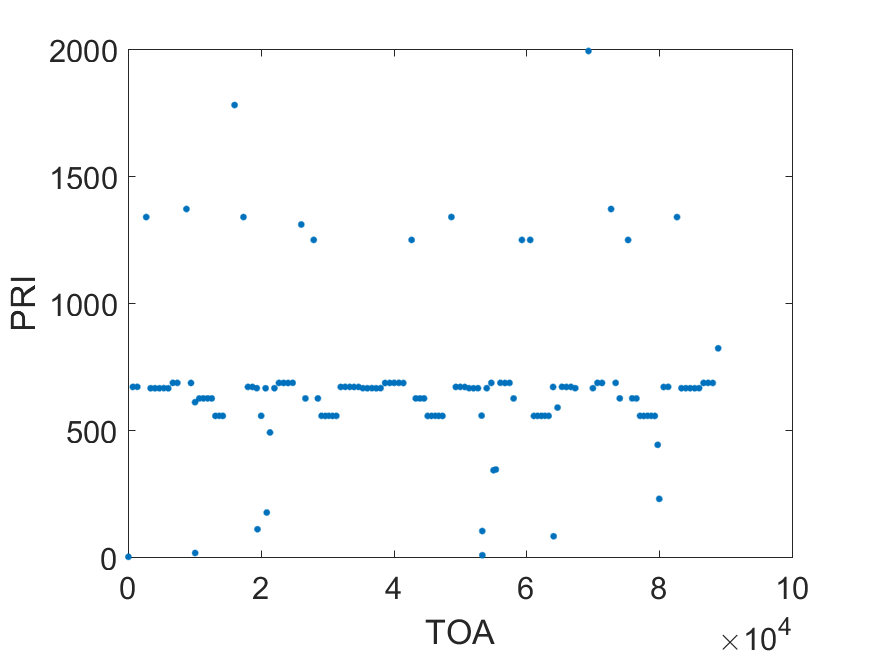}%
\label{exp2_3}}
\hfill
\subfloat[PW features of target 2.]{\includegraphics[width=1.6in]{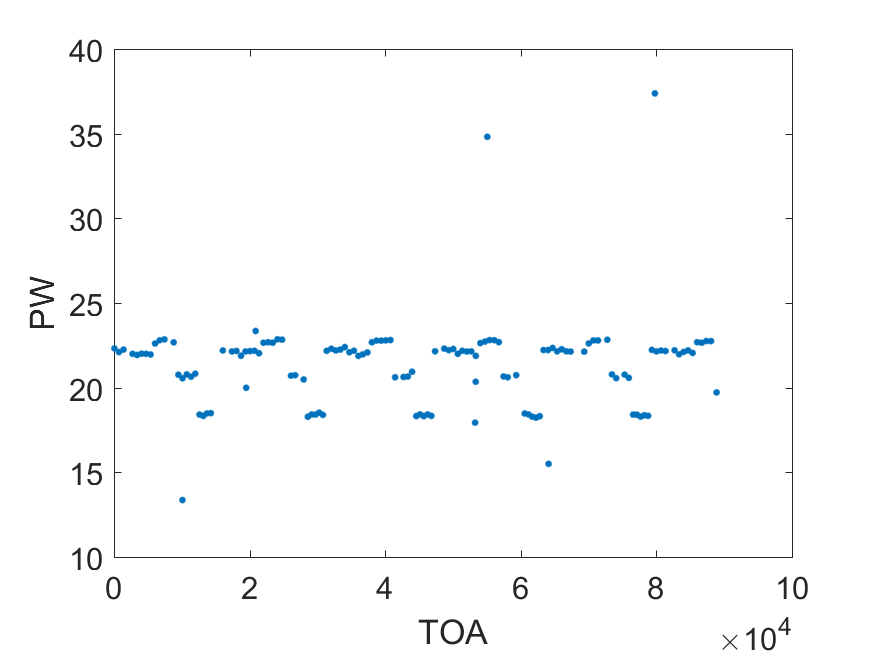}%
\label{exp2_4}}
\hfill
\subfloat[PRI features of target 3.]{\includegraphics[width=1.6in]{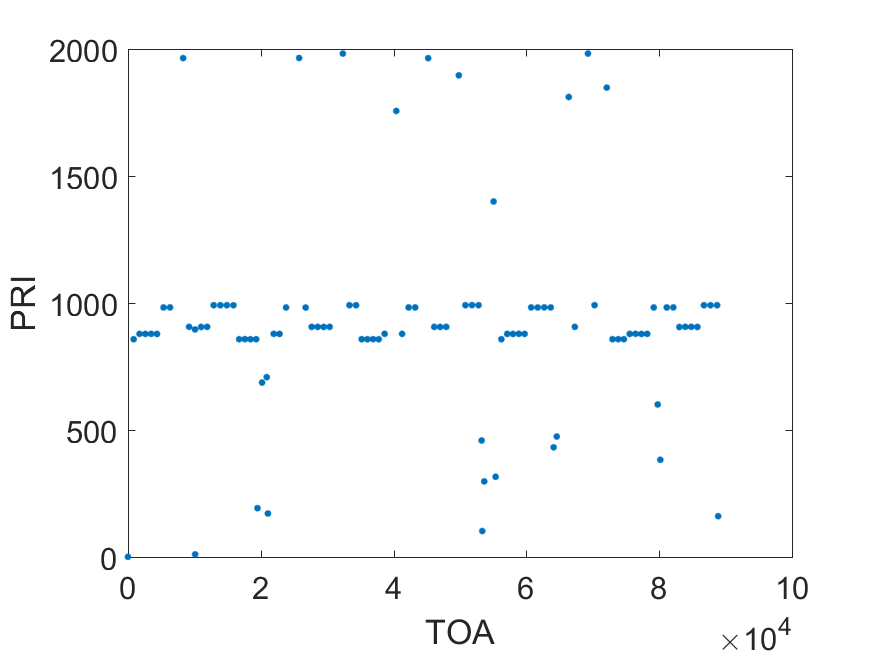}%
\label{exp2_5}}
\hfill
\subfloat[PW features of target 3.]{\includegraphics[width=1.6in]{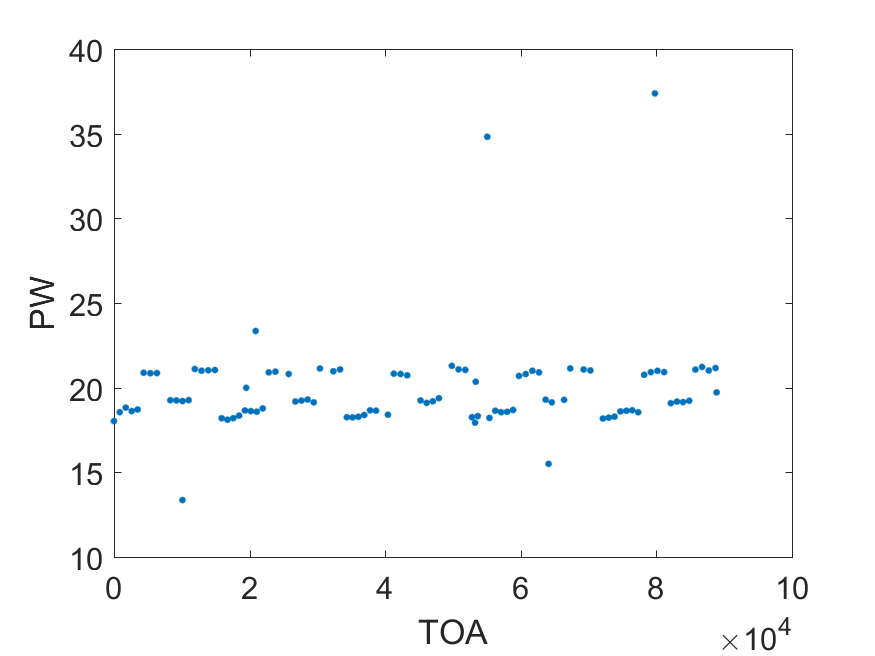}%
\label{exp2_6}}
\hfill
\subfloat[DTOA of interleaved pulse stream in Experiment 2.]{\includegraphics[width=1.6in]{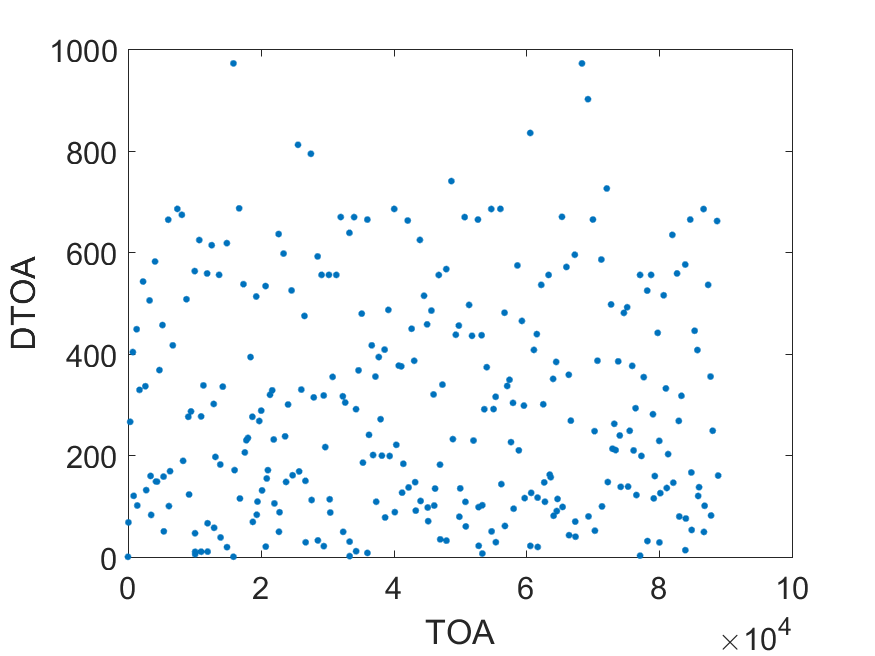}%
\label{exp2_7}}
\hfil
\subfloat[PW of interleaved pulse stream in Experiment 2.]{\includegraphics[width=1.6in]{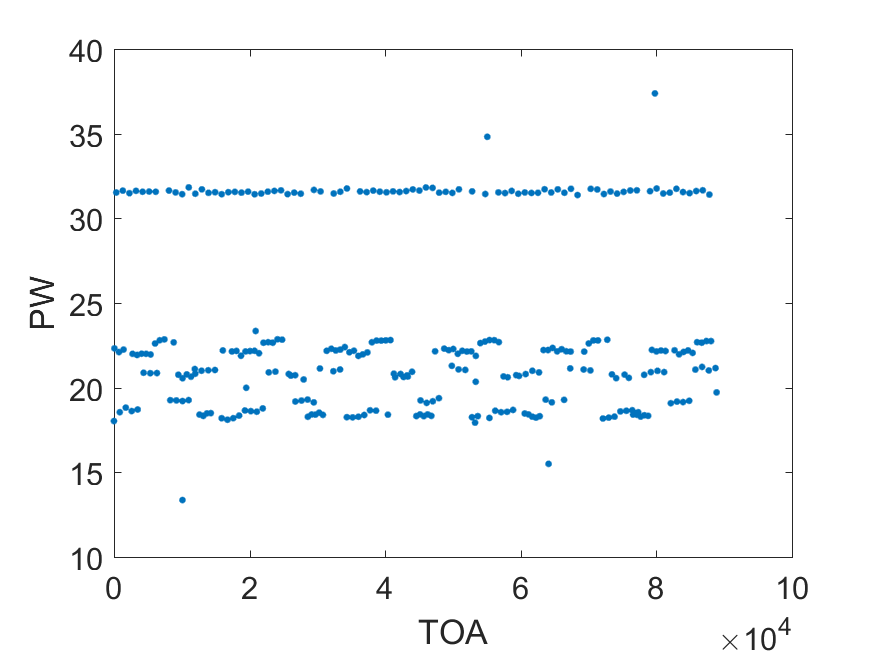}%
\label{exp2_8}}
\caption{Signal features of target radars in Experiment 2.}
\label{exp2}
\end{figure}

\begin{figure}[!t]
\centering
\subfloat[PRI features of target 1.]{\includegraphics[width=1.6in]{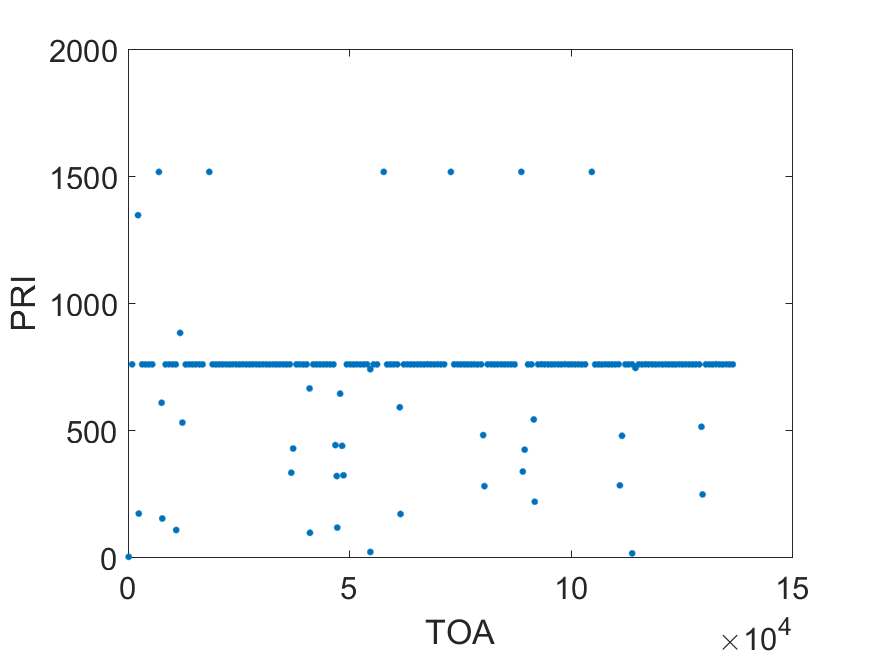}%
\label{exp3_1}}
\hfill
\subfloat[RF features of target 1.]{\includegraphics[width=1.6in]{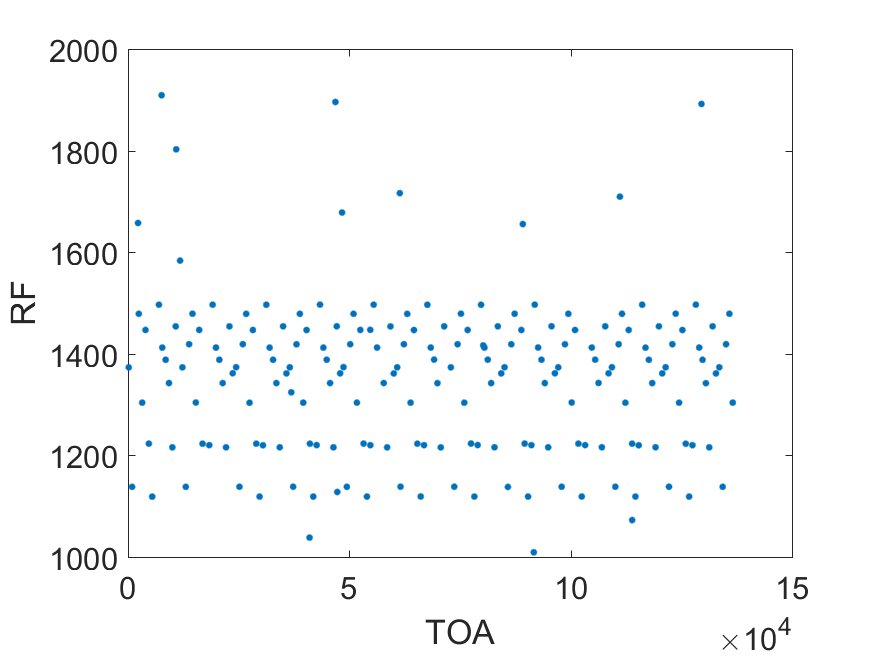}%
\label{exp3_2}}
\hfill
\subfloat[PRI features of target 2.]{\includegraphics[width=1.6in]{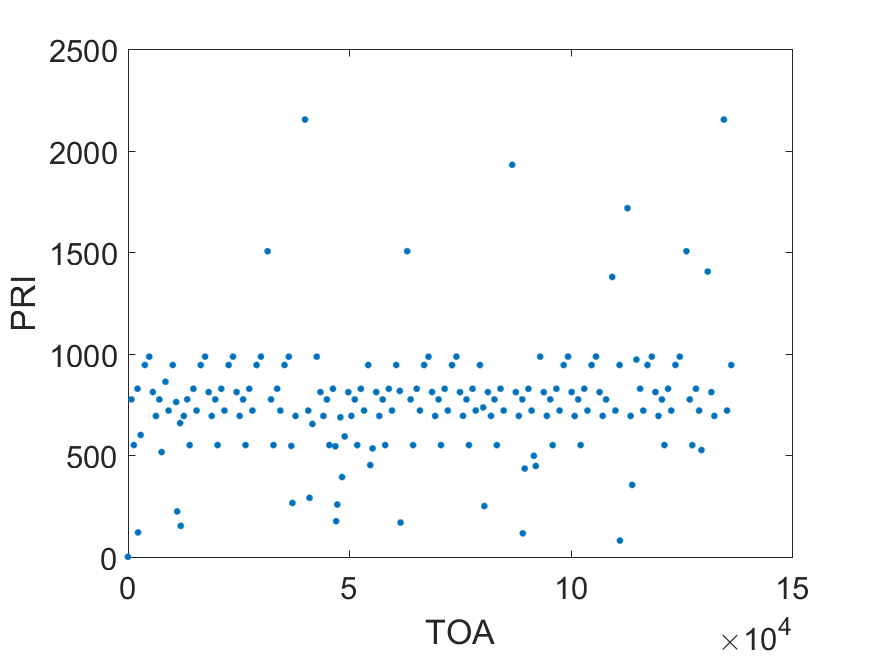}%
\label{exp3_3}}
\hfill
\subfloat[RF features of target 2.]{\includegraphics[width=1.6in]{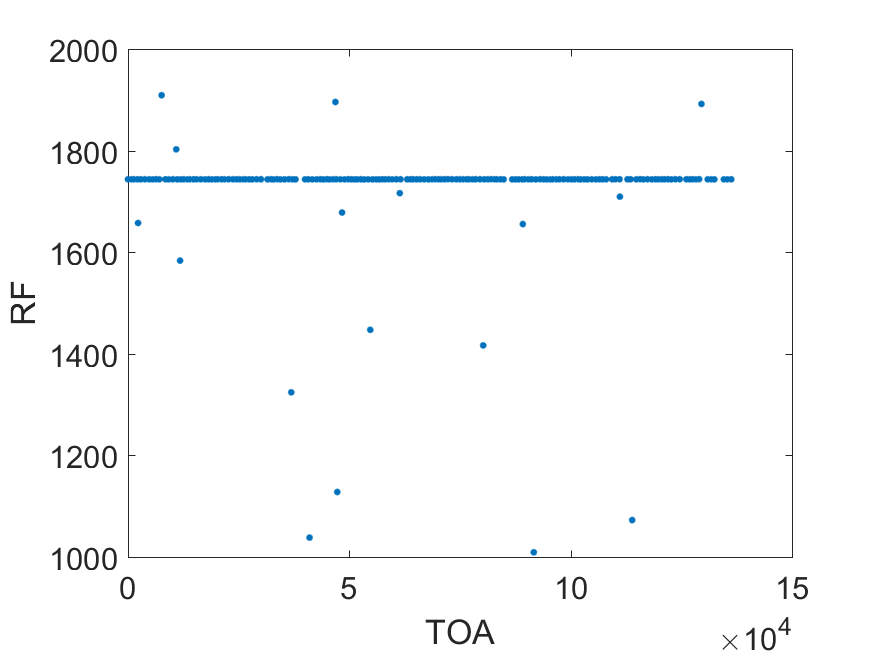}%
\label{exp3_4}}
\hfill
\subfloat[PRI features of target 3.]{\includegraphics[width=1.6in]{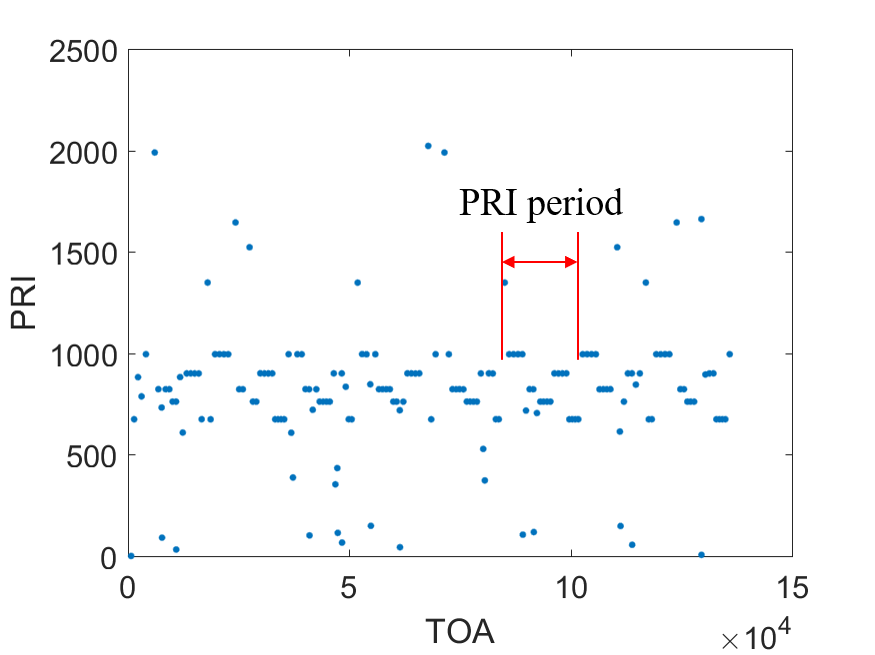}%
\label{exp3_5}}
\hfill
\subfloat[RF features of target 3.]{\includegraphics[width=1.6in]{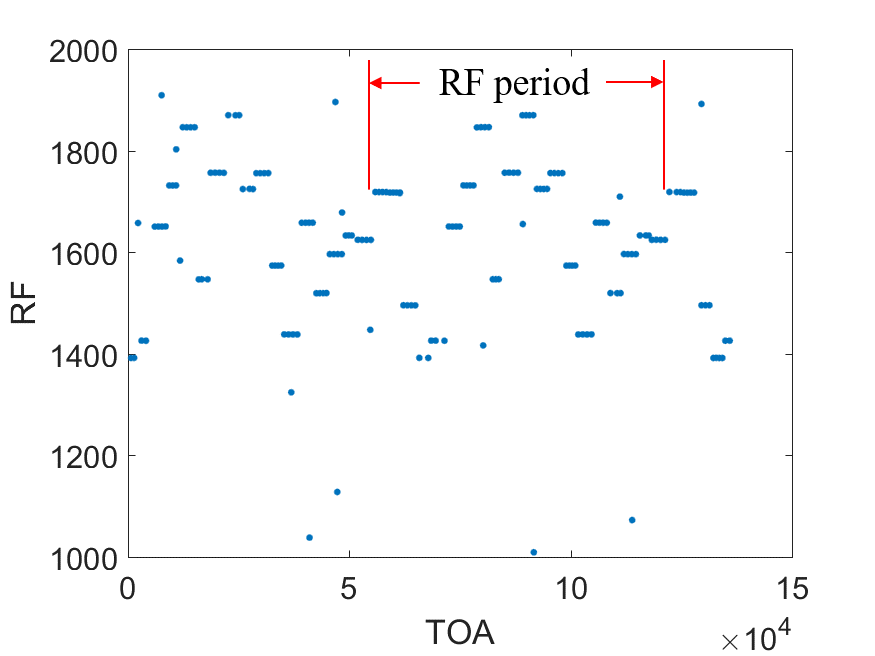}%
\label{exp3_6}}
\hfill
\subfloat[DTOA of interleaved pulse stream in Experiment 3.]{\includegraphics[width=1.6in]{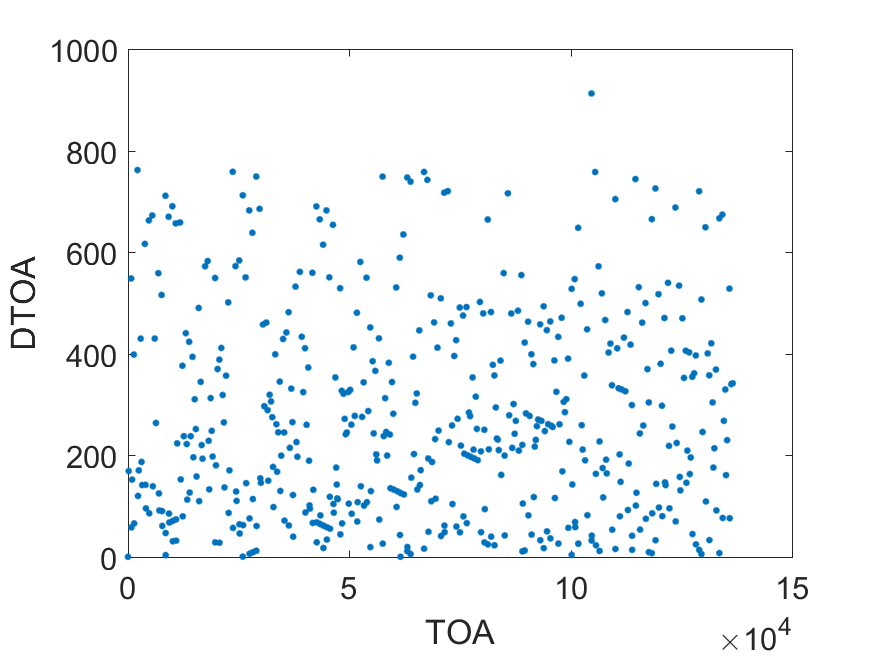}%
\label{exp3_7}}
\hfil
\subfloat[RF of interleaved pulse stream in Experiment 3.]{\includegraphics[width=1.6in]{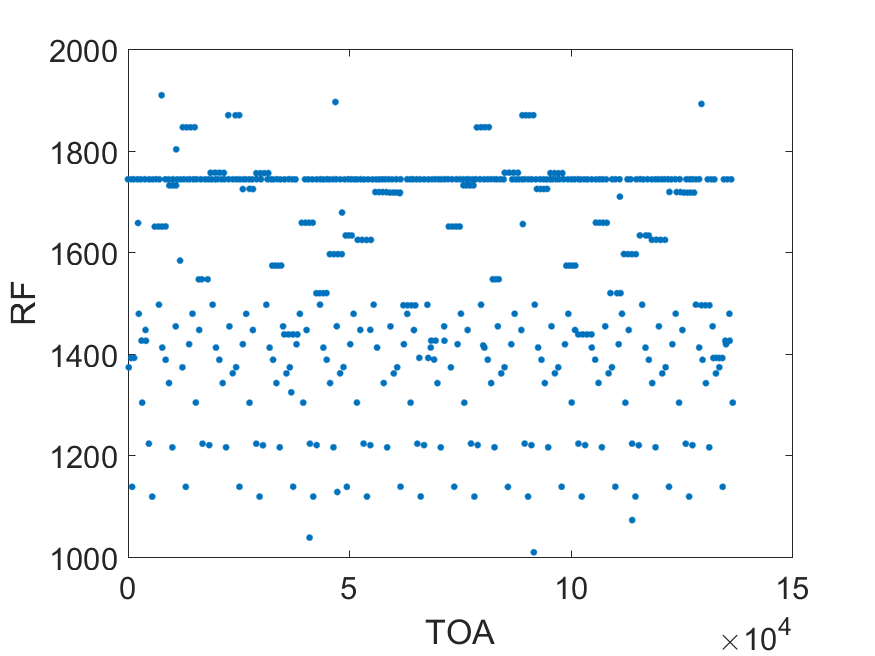}%
\label{exp3_8}}
\caption{Signal features of target radars in Experiment 3.}
\label{exp3}
\end{figure}

\begin{figure}[!t]
\centering
\subfloat[PRI features of target 1.]{\includegraphics[width=1.6in]{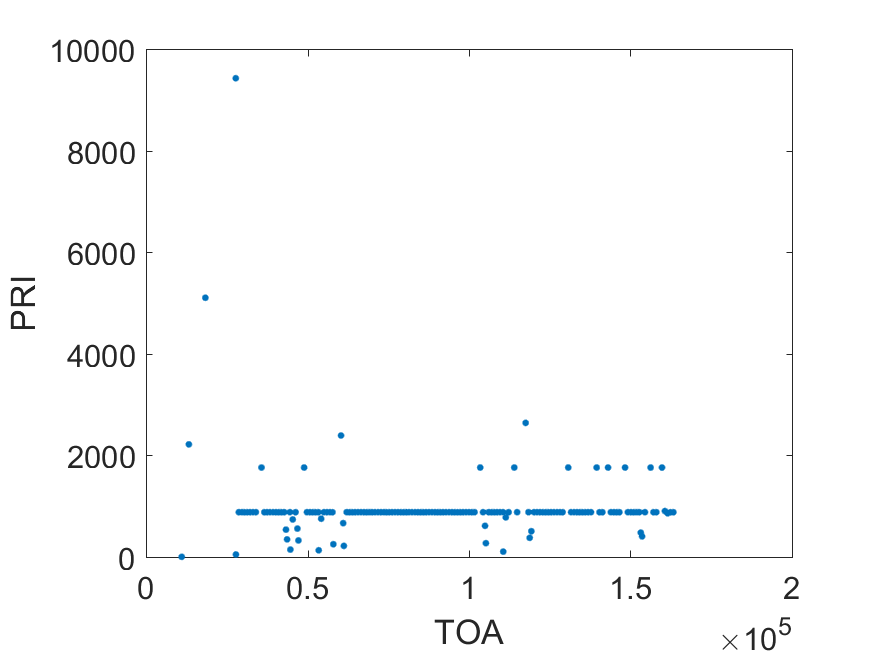}%
\label{exp4_1}}
\hfill
\subfloat[PA features of target 1.]{\includegraphics[width=1.6in]{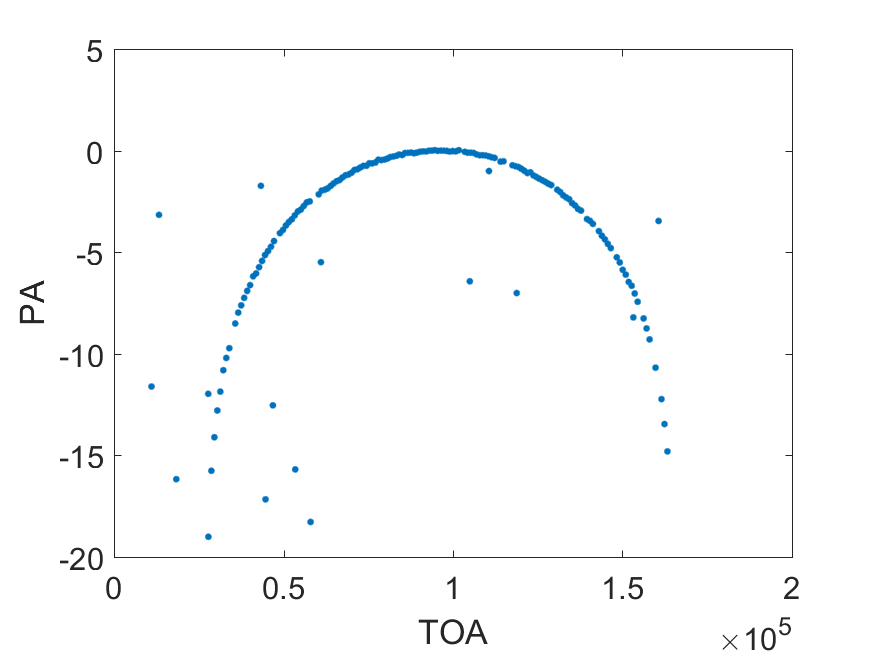}%
\label{exp4_2}}
\hfill
\subfloat[PRI features of target 2.]{\includegraphics[width=1.6in]{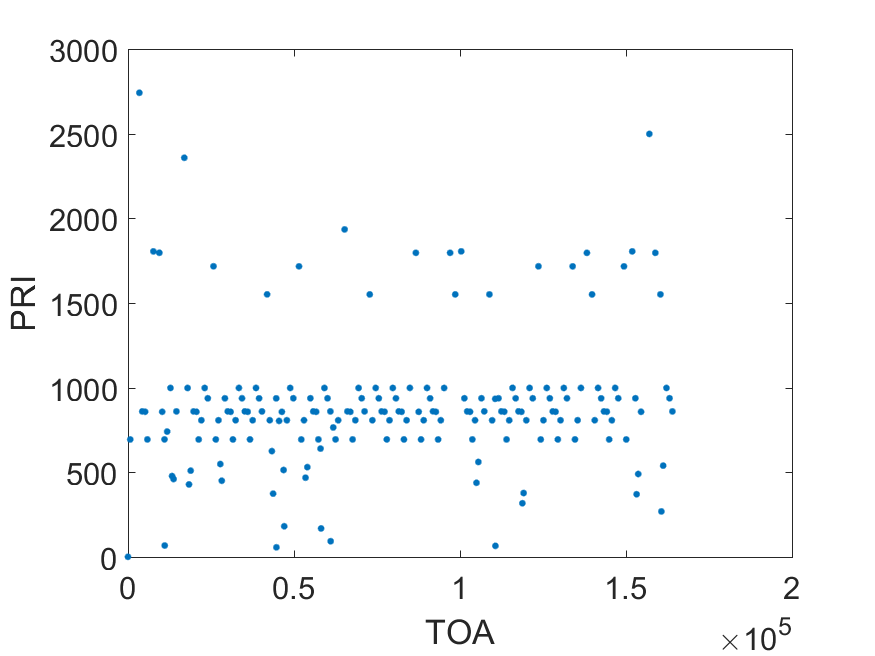}%
\label{exp4_3}}
\hfill
\subfloat[PA features of target 2.]{\includegraphics[width=1.6in]{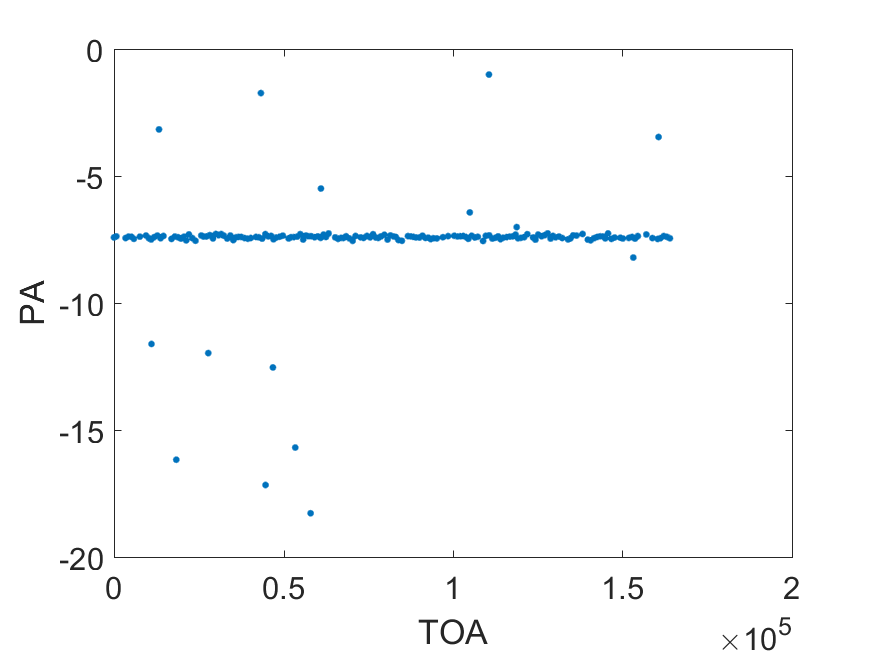}%
\label{exp4_4}}
\hfill
\subfloat[PRI features of target 3.]{\includegraphics[width=1.6in]{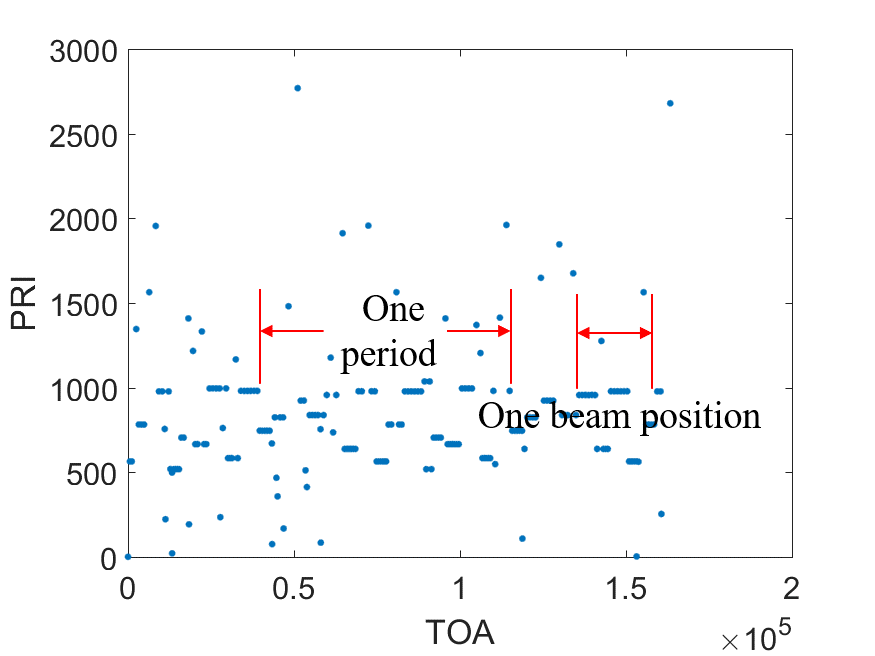}%
\label{exp4_5}}
\hfill
\subfloat[PA features of target 3.]{\includegraphics[width=1.6in]{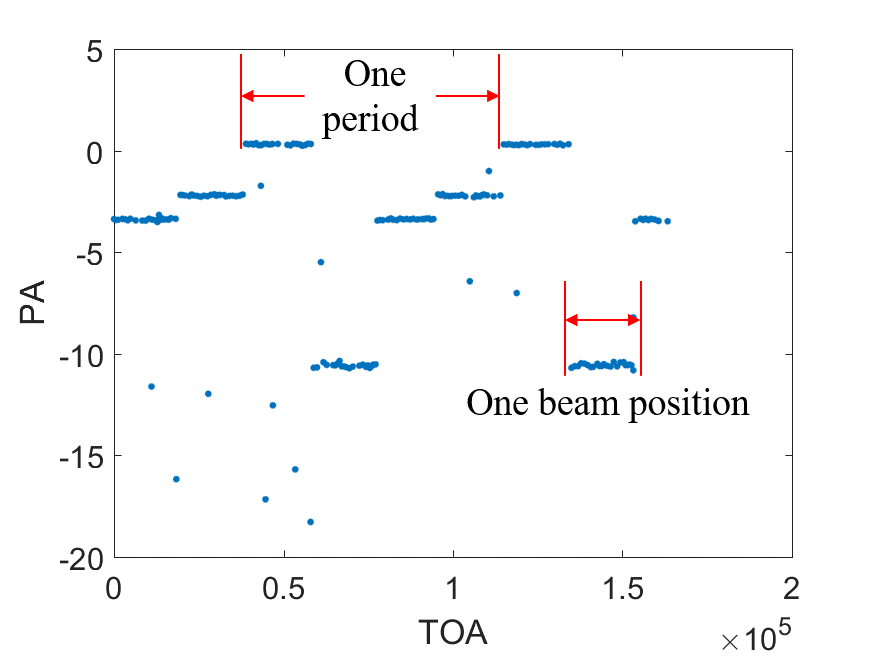}%
\label{exp4_6}}
\hfill
\subfloat[DTOA of interleaved pulse stream in Experiment 4.]{\includegraphics[width=1.6in]{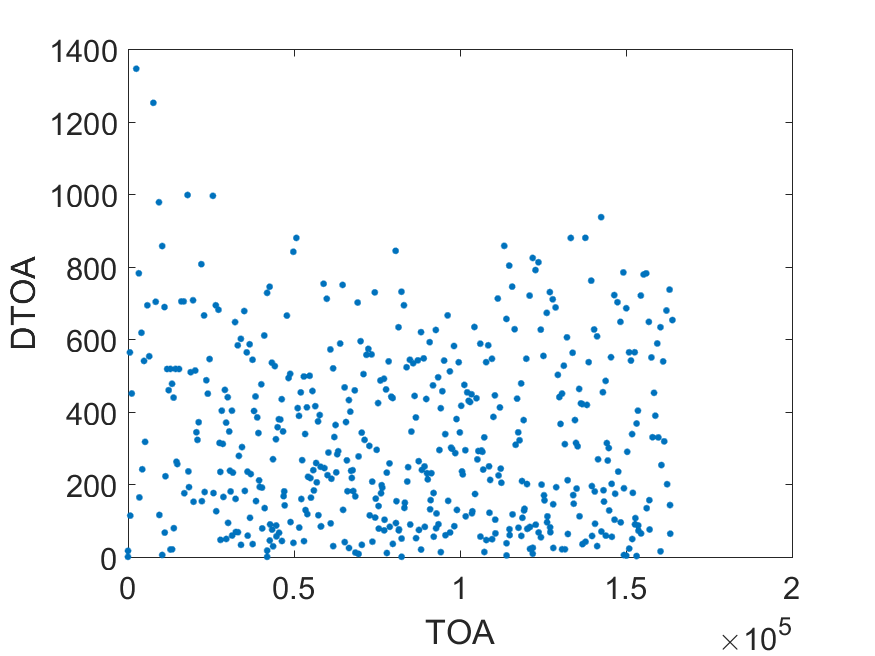}%
\label{exp4_7}}
\hfil
\subfloat[PA of interleaved pulse stream in Experiment 4.]{\includegraphics[width=1.6in]{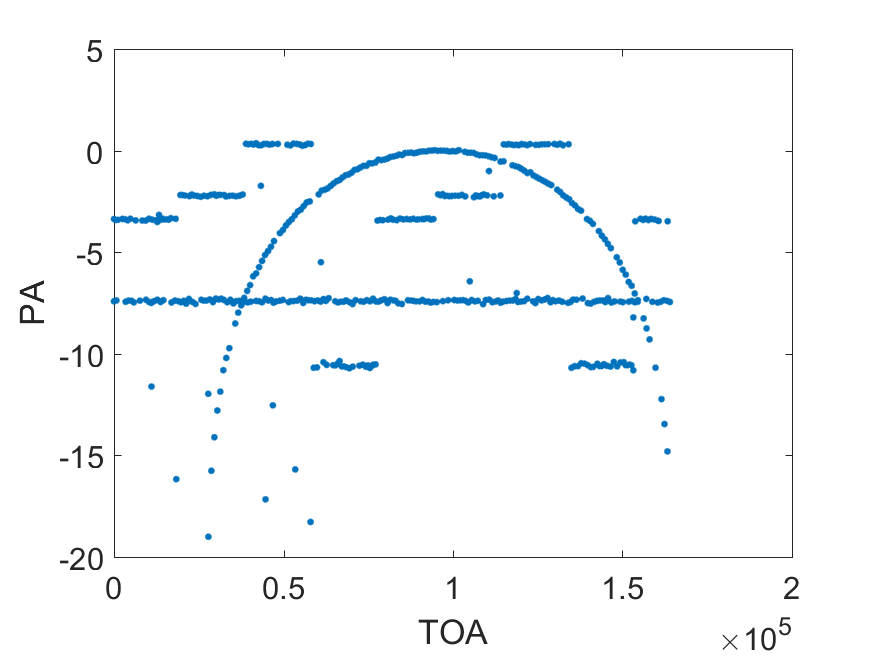}%
\label{exp4_8}}
\caption{Signal features of target radars in Experiment 4.}
\label{exp4}
\end{figure}

In this section, four experiments are conducted to verify the feasibility and efficiency of the proposed deinterleaving method.
\textbf{In the four experiments}, the values of RF, PW, and PA of target radars vary, and the value range overlap. \textbf{The existing multiparameter-based deinterleaving methods are not able to take advantage of these parameters}.
\textbf{In experiment 1}, the target radars have the same PRI modulation modes and value ranges. 
\textbf{The SSD-PRI method is not able to deinterleave the pulses in this case}.
The experiments are used to verify these problems.

1) Experiments 1–4 can verify that, the SSD-Multipara method exhibits much higher accuracy than the PRI-based method.

2) Experiment 1 can verify that, in some cases, the SSD-Multipara can deinterleave pulses with high accuracy, when the target radars have the same PRI features and the pulses deinterleaving cannot be achieved by SSD-PRI method.

3) Experiments 1–4 can verify that, the SSD-Multipara method can adapt to radar signals with variable parameters and cases where the parameter values of different targets overlap, while the existing multiparameter-based deinterleaving methods cannot adapt to these situations.

4) Experiments 1–4 can verify that, the RF, PW, and PA can be used together with the PRI for deinterleaving in one step. 

The ${\rho _l}$ and ${\rho _n}$ of each sample of the training data are randomly selected within the 0–0.5 range. Figs. \ref{exp1}–\ref{exp4} show the target features obtained from experiments 1–4. In these figures, ${\rho _l} = 0.1$ and ${\rho _n} = 0.1$ were set.

Since \textbf{the existing multiparameter-based deinterleaving methods are incapable of solving the deinterleaving problems set in the simulation}, three PRI-based deinterleaving methods are selected as the control experiment: SSD-PRI, SDIF, and PRI-Tran.

\subsubsection{Experiment 1–Deinterleaving radar signals with DTOA/PW}
The target settings are shown in Table \ref{tab_exp1}, and the target features are shown in Fig. \ref{exp1}. It can be observed that the PRI modulation mode and the PW variation mode of the three targets are the same, and the value ranges of PRI and PW overlap. 

The difference between the targets is the DC. The SSD-PRI method can not adapt to this case, and is not chosen as the control experiment. 

In this case, the existing multiparameter-based deinterleaving methods cannot cluster the pulses by PW. The PW and PRI
belonging to the same target are coupled.

\begin{table*}[]
\centering
\caption{Target settings in Experiment 1}
\label{tab_exp1}
\resizebox{6.5in}{!}{
\begin{tabular}{|c|llc|}
\hline
\textbf{Category} &
  \multicolumn{1}{c|}{\textbf{PRI features}} &
  \multicolumn{1}{c|}{\textbf{PW features}} &
  \textbf{Number of targets} \\ \hline
1 &
  \multicolumn{1}{l|}{\begin{tabular}[c]{@{}l@{}}Modulation mode:   D\&S\\ Value range: 500–1000\\ $J$ (number of pulses in each group): 4–6\\ $K$ (number of pulse groups in one   period): 4–6\end{tabular}} &
  \multicolumn{1}{l|}{\begin{tabular}[c]{@{}l@{}}Variation mode: D\&S   \\ Value: $PRI*DC$\\ $DC = 0.03$\end{tabular}} &
  1 \\ \hline
2 &
  \multicolumn{1}{l|}{Same with target 1} &
  \multicolumn{1}{l|}{\begin{tabular}[c]{@{}l@{}}Variation mode: D\&S\\ Value: $PRI*DC$\\ $DC = 0.04$\end{tabular}} &
  1 \\ \hline
3 &
  \multicolumn{1}{l|}{Same with target 1} &
  \multicolumn{1}{l|}{\begin{tabular}[c]{@{}l@{}}Variation mode: D\&S\\ Value: $PRI*DC$\\ $DC = 0.05$\end{tabular}} &
  1 \\ \hline
4 &
  \multicolumn{3}{l|}{\begin{tabular}[c]{@{}l@{}}Random noise pulses, $PW = c*d$,   $c \sim U(500,1000)$, $d \sim U(0.01,0.05)$. For each   pulse, \\  $c$ and $d$ obtain random values.\end{tabular}} \\ \hline
\end{tabular}
}
\end{table*}

\subsubsection{Experiment 2–Deinterleaving radar signals with DTOA/PW}
The target settings are shown in Table \ref{tab_exp2}, and the target features are shown in Fig. \ref{exp2}. Target 2 and Target 3 adopt the same PRI modulation mode, but the PRI values do not overlap. Target 1 adopts a different PRI modulation mode, but the PRI value range overlap with the other two targets. The PW variation modes of targets 2 and 3 are the same. The PW value ranges of the three targets overlap. 

In this case, the existing multiparameter-based deinterleaving methods cannot cluster the pulses by PW. The PW and PRI belonging to the same target are coupled.

\begin{table*}[]
\centering
\caption{Target settings in Experiment 2}
\label{tab_exp2}
\resizebox{6.5in}{!}{
\begin{tabular}{|c|llc|}
\hline
\textbf{Category} &
  \multicolumn{1}{c|}{\textbf{PRI features}} &
  \multicolumn{1}{c|}{\textbf{PW features}} &
  \textbf{Number of targets} \\ \hline
1 &
  \multicolumn{1}{l|}{\begin{tabular}[c]{@{}l@{}}Modulation mode: constant   \\ Value range: 500–1000\end{tabular}} &
  \multicolumn{1}{l|}{\begin{tabular}[c]{@{}l@{}}Variation mode: constant\\ Value: $PRI*DC$\\ $DC \sim U(0.01,0.05)$  \\During a signal transmission, \\the radar DC remains unchanged. \end{tabular}} &
  1 \\ \hline
2 &
  \multicolumn{1}{l|}{\begin{tabular}[c]{@{}l@{}}Modulation mode:   D\&S   \\ Value range: 500–750   \\ $J$ (number of pulses in each group): 4–6\\ $K$ (number of pulse groups in one   period): 4–6\end{tabular}} &
  \multicolumn{1}{l|}{\begin{tabular}[c]{@{}l@{}}Variation mode: D\&S   \\ Value: $PRI*DC$   \\ $DC \sim U(0.01,0.05)$  \\During a signal transmission, \\the radar DC remains unchanged. \end{tabular}} &
  1 \\ \hline
3 &
  \multicolumn{1}{l|}{\begin{tabular}[c]{@{}l@{}}Modulation mode:   D\&S   \\ Value range: 750–1000   \\ $J$: 4–6   \\ $K$: 4–6\end{tabular}} &
  \multicolumn{1}{l|}{\begin{tabular}[c]{@{}l@{}}Variation mode: D\&S   \\ Value: $PRI*DC$   \\ $DC \sim U(0.01,0.05)$  \\During a signal transmission, \\the radar DC remains unchanged. \end{tabular}} &
  1 \\ \hline
4 &
  \multicolumn{3}{l|}{\begin{tabular}[c]{@{}l@{}}Random noise pulses, $PW = c*d$,   $c \sim U(500,1000)$, $d \sim U(0.01,0.05)$. \\For each   pulse, $c$ and $d$ obtain random values.\end{tabular}} \\ \hline
\end{tabular}
}
\end{table*}

\subsubsection{Experiment 3–Deinterleaving radar signals with DTOA/RF}
The target settings are shown in Table \ref{tab_exp3}, and the target features are shown in Fig. \ref{exp3}. The PRI modulation modes and the RF variation modes of the three targets are different, but the value ranges overlap.

In this case, the existing multiparameter-based deinterleaving methods cannot cluster the pulses by RF. The RF and PRI
belonging to the same target are coupled.

\begin{table*}[]
\centering
\caption{Target settings in Experiment 3}
\label{tab_exp3}
\resizebox{6.5in}{!}{
\begin{tabular}{|c|llc|}
\hline
\textbf{Category} &
  \multicolumn{1}{c|}{\textbf{PRI features}} &
  \multicolumn{1}{c|}{\textbf{RF features}} &
  \textbf{Number of targets} \\ \hline
1 &
  \multicolumn{1}{l|}{\begin{tabular}[c]{@{}l@{}}Modulation mode: constant\\ Value range: 500–1000\end{tabular}} &
  \multicolumn{1}{l|}{\begin{tabular}[c]{@{}l@{}}Variation mode: agile RF between   pulses\\ Value range: 1000–2000\\ $M$ (number of RF values in one period): 10–20\end{tabular}} &
  1 \\ \hline
2 &
  \multicolumn{1}{l|}{\begin{tabular}[c]{@{}l@{}}Modulation mode: staggered\\ Value range: 500–1000\\ $L$ (number of PRI values in \\ one period):3–10\end{tabular}} &
  \multicolumn{1}{l|}{\begin{tabular}[c]{@{}l@{}}Variation mode: constant\\ Value range: 1000–2000\end{tabular}} &
  1 \\ \hline
3 &
  \multicolumn{1}{l|}{\begin{tabular}[c]{@{}l@{}}Modulation mode:   D\&S\\ Value range: 500–1000\\ $J$ (number of pulses in each group): 4–6\\ $K$ (number of pulse groups in one   period): 4–6\end{tabular}} &
  \multicolumn{1}{l|}{\begin{tabular}[c]{@{}l@{}}Variation mode: agile RF between pulse   groups\\ Value range: 1000–2000\\ $O$ (number of pulses in each group): $O=J$\\ $P$ (number of RF groups in one period): 10–20\\RF varies synchronously with PRI, but the\\ PRI period is different from the RF period,\\as shown in Figs. \ref{exp3_5} and \ref{exp3_6}.\end{tabular}} &
  1 \\ \hline
4 &
  \multicolumn{3}{l|}{Random noise pulses, $RF \sim U(1000,2000$).} \\ \hline
\end{tabular}
}
\end{table*}

\subsubsection{Experiment 4–Deinterleaving radar signals with DTOA/PA}
The target settings are shown in Table \ref{tab_exp4}, and the target features are shown in Fig. \ref{exp4}. The PRI modulation modes and PA variation modes of the three targets are different, but the value ranges overlap.

In this case, the existing multiparameter-based deinterleaving methods cannot cluster the pulses by PA. The PA and PRI
belonging to the same target are coupled.

\begin{table*}[]
\centering
\caption{Target settings in Experiment 4}
\label{tab_exp4}
\resizebox{6.5in}{!}{
\begin{tabular}{|c|llc|}
\hline
\textbf{Category} &
  \multicolumn{1}{c|}{\textbf{PRI features}} &
  \multicolumn{1}{c|}{\textbf{PA features}} &
  \textbf{Number of targets} \\ \hline
1 &
  \multicolumn{1}{l|}{\begin{tabular}[c]{@{}l@{}}Modulation mode: constant   \\ Value range: 500–1000\end{tabular}} &
  \multicolumn{1}{l|}{\begin{tabular}[c]{@{}l@{}}Variation mode: radar mechanical   scanning   \\ Value range: 1000–2000   \\ $a$ (beamwidth coefficient): 11.5   \\ Scanning speed: 10 seconds per   round\end{tabular}} &
  1 \\ \hline
2 &
  \multicolumn{1}{l|}{\begin{tabular}[c]{@{}l@{}}Modulation mode: staggered   \\ Value range: 500–1000   \\ $L$: 3–10\end{tabular}} &
  \multicolumn{1}{l|}{Variation mode: radar non-scanning} &
  1 \\ \hline
3 &
  \multicolumn{1}{l|}{\begin{tabular}[c]{@{}l@{}}Modulation mode:   D\&S   \\ Value range: 500–1000   \\ $K$ (number of pulse groups in one period): $S*T$, i.e., 16   \\ $J$(number of pulses in each group): 4–6\end{tabular}} &
  \multicolumn{1}{l|}{\begin{tabular}[c]{@{}l@{}}Variation mode: radar phase scanning   \\ $S$ (number of beam   positions in one period): 4   \\ $T$ (number of pulse groups in a beam position): 4   \\ $R$ (number of pulses in   each beam position): $K*J$\\The scanning period is the same as the PRI period, \\with $S$ beam positions per periode, $T$ pulse groups\\ per beam position, and $J$ pulses per pulse group. The \\number of pulses in each period is \\$K*J=S*T*J$, \\as shown in Figs. \ref{exp4_5} and \ref{exp4_6}.\end{tabular}} &
  1 \\ \hline
4 &
  \multicolumn{3}{l|}{\begin{tabular}[c]{@{}l@{}}Random noise pulses, PA is   randomly selected between the maximum PA and minimum PA of the target\\  radar   signals.\end{tabular}} \\ \hline
\end{tabular}
}
\end{table*}



\subsection{Results}
The overall performance on the test set (produced under the same conditions as the training set) in the four experiments is listed in Table \ref{tab_res}.

\begin{table}[]
\centering
\caption{Overall performance on the test set in the four experiments}
\label{tab_res}
\resizebox{3in}{!}{
\begin{tabular}{|p{1.5cm}<{\centering}|p{3cm}<{\centering}|p{1.5cm}<{\centering}|}
\hline
\textbf{Experiment} & \textbf{Method} & \textbf{Accuracy} \\ \hline
\multirow{3}{*}{1}  & SSD-Multipara   & 99.4              \\ \cline{2-3} 
                    & SDIF            & 63.01             \\ \cline{2-3} 
                    & PRI-Tran        & 62.94             \\ \hline
\multirow{4}{*}{2}  & SSD-Multipara   & 99.8              \\ \cline{2-3} 
                    & SSD-PRI         & 93.7              \\ \cline{2-3} 
                    & SDIF            & 89.56             \\ \cline{2-3} 
                    & PRI-Tran        & 70.64             \\ \hline
\multirow{4}{*}{3}  & SSD-Multipara   & 96.2              \\ \cline{2-3} 
                    & SSD-PRI         & 87                \\ \cline{2-3} 
                    & SDIF            & 90.52             \\ \cline{2-3} 
                    & PRI-Tran        & 62.44             \\ \hline
\multirow{4}{*}{4}  & SSD-Multipara   & 99.6              \\ \cline{2-3} 
                    & SSD-PRI         & 90.6              \\ \cline{2-3} 
                    & SDIF            & 65.77             \\ \cline{2-3} 
                    & PRI-Tran        & 22.41             \\ \hline
\end{tabular}
}
\end{table}

\begin{figure*}[!t]
\centering
\subfloat[]{\includegraphics[width=2.3in]{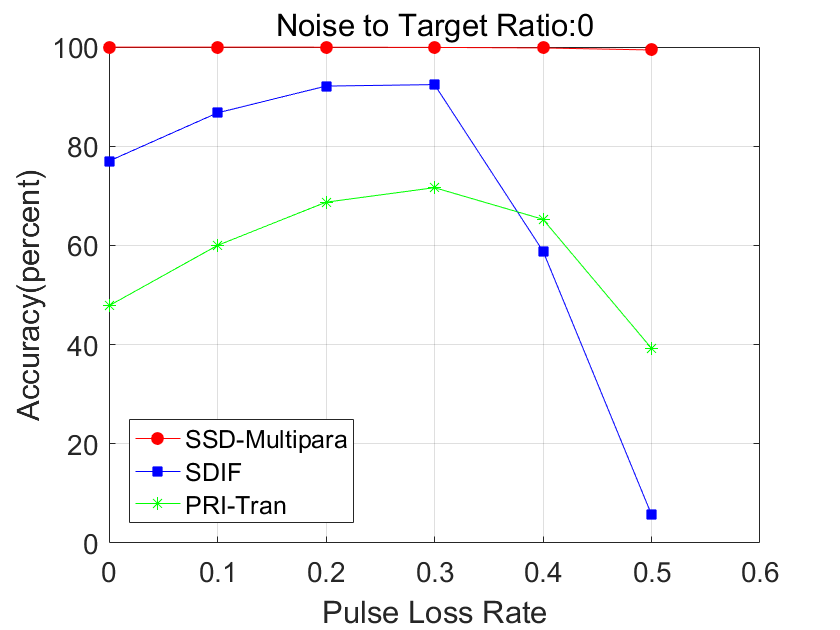}%
\label{exp1a}}
\hfil
\subfloat[]{\includegraphics[width=2.3in]{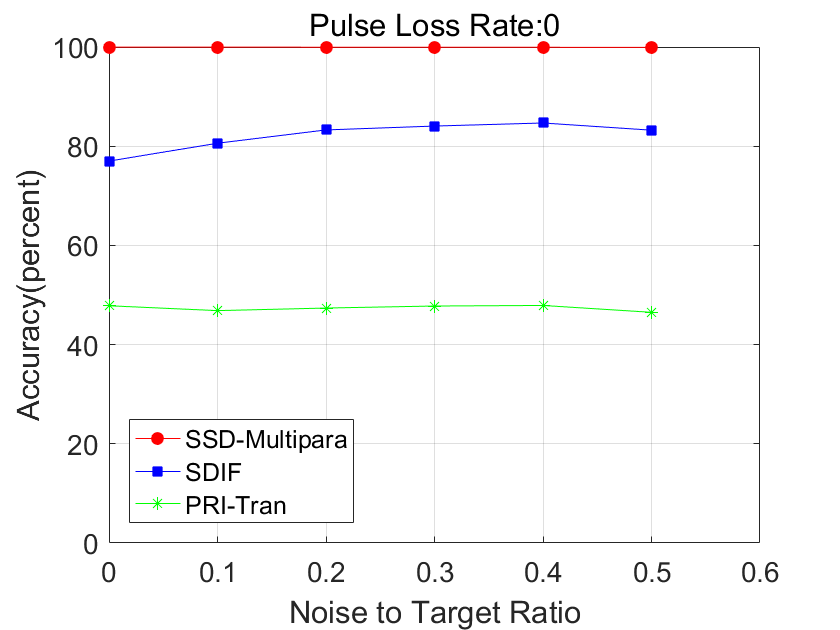}%
\label{exp1b}}
\hfil
\subfloat[]{\includegraphics[width=2.3in]{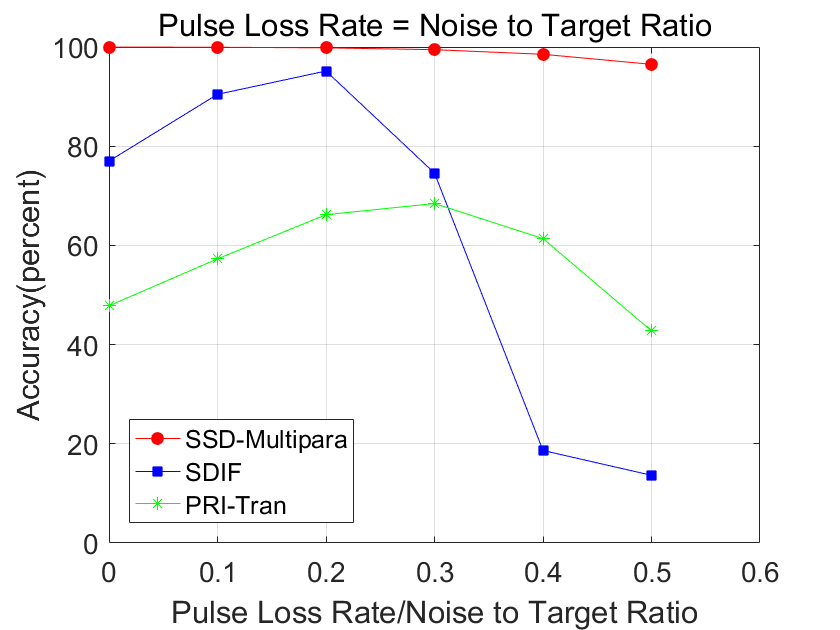}%
\label{exp1c}}
\caption{Test results of Experiment 1.}
\label{exp1_res}
\end{figure*}

\begin{figure*}[!t]
\centering
\subfloat[]{\includegraphics[width=2.3in]{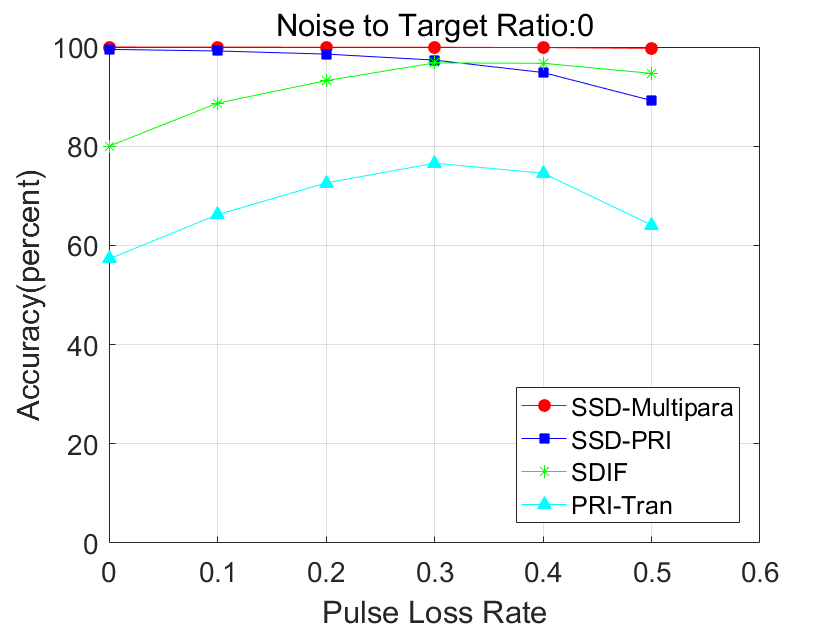}%
\label{exp2a}}
\hfil
\subfloat[]{\includegraphics[width=2.3in]{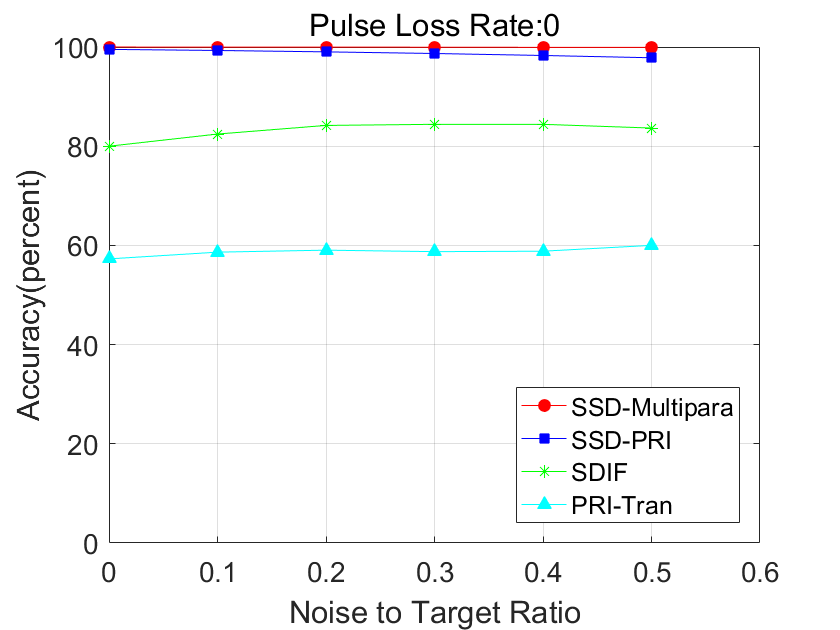}%
\label{exp2b}}
\hfil
\subfloat[]{\includegraphics[width=2.3in]{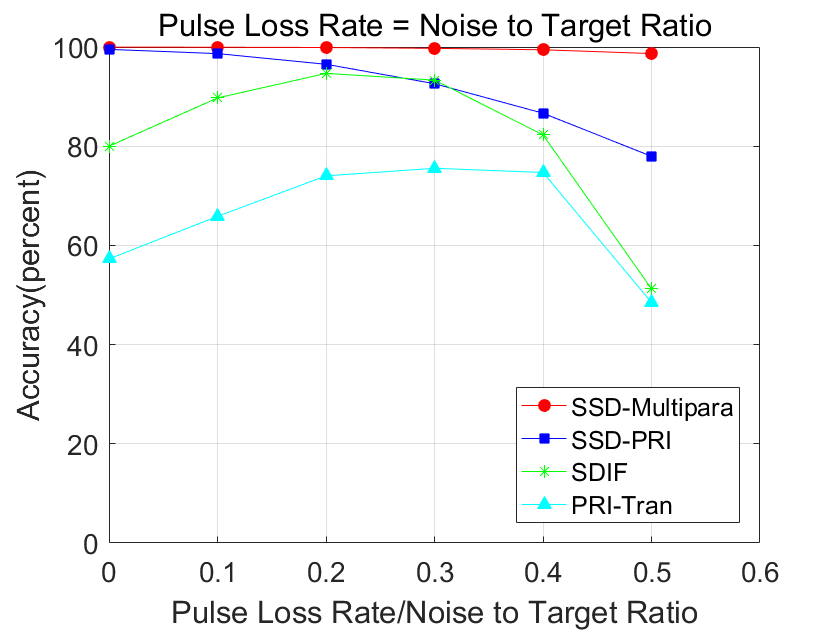}%
\label{exp2c}}
\caption{Test results of Experiment 2.}
\label{exp2_res}
\end{figure*}

\begin{figure*}[!t]
\centering
\subfloat[]{\includegraphics[width=2.3in]{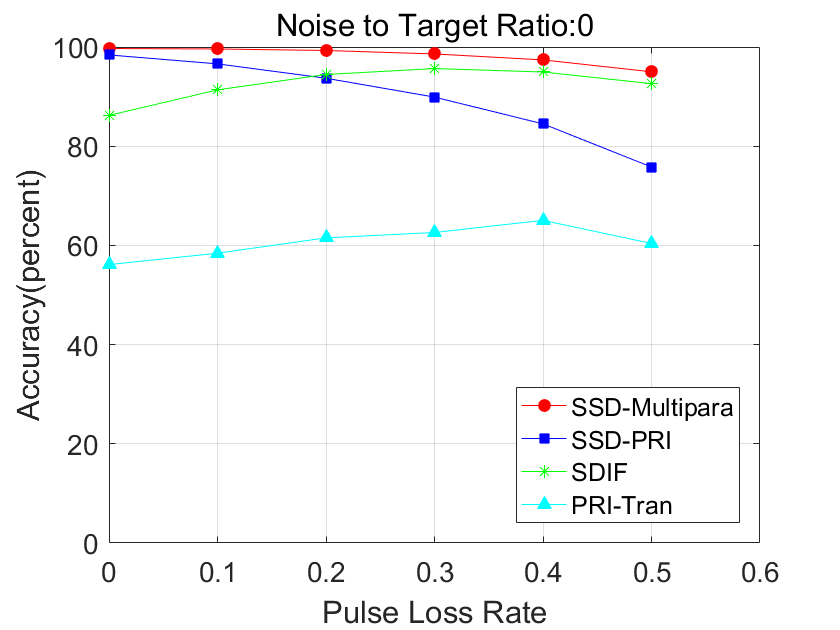}%
\label{exp3a}}
\hfil
\subfloat[]{\includegraphics[width=2.3in]{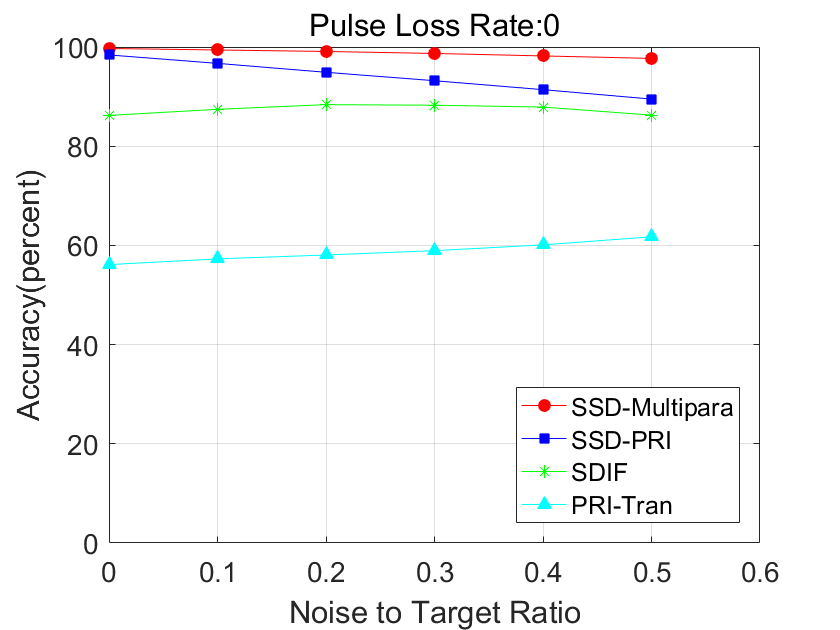}%
\label{exp3b}}
\hfil
\subfloat[]{\includegraphics[width=2.3in]{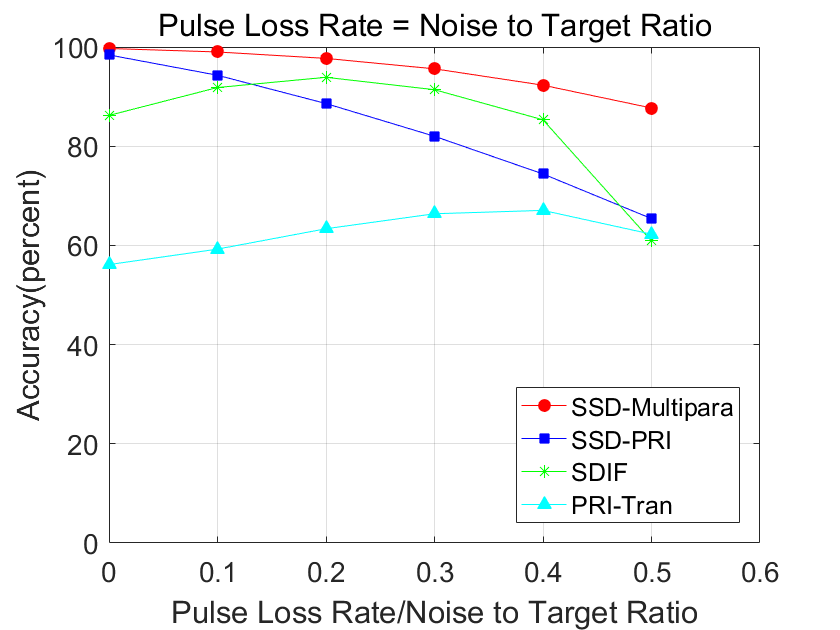}%
\label{exp3c}}
\caption{Test results of Experiment 3.}
\label{exp3_res}
\end{figure*}

\begin{figure*}[!t]
\centering
\subfloat[]{\includegraphics[width=2.3in]{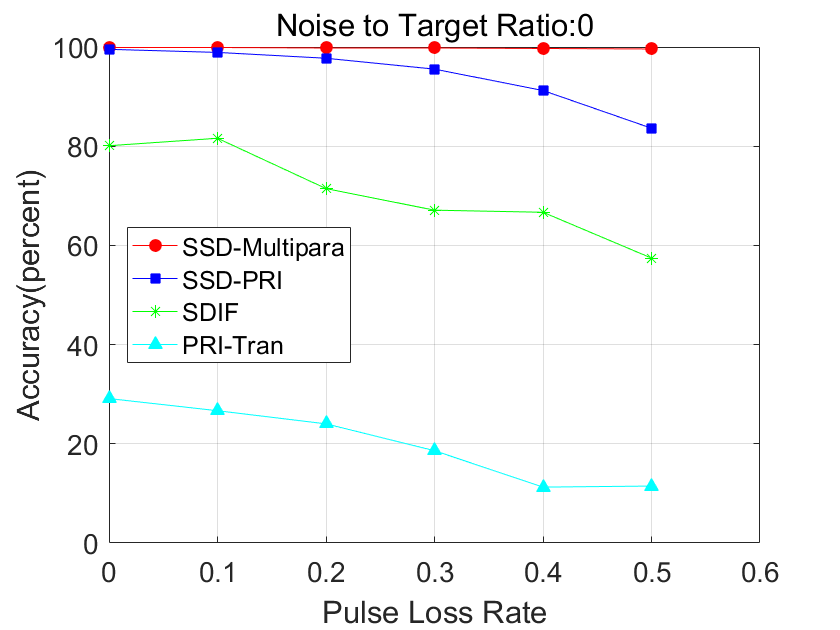}%
\label{exp4a}}
\hfil
\subfloat[]{\includegraphics[width=2.3in]{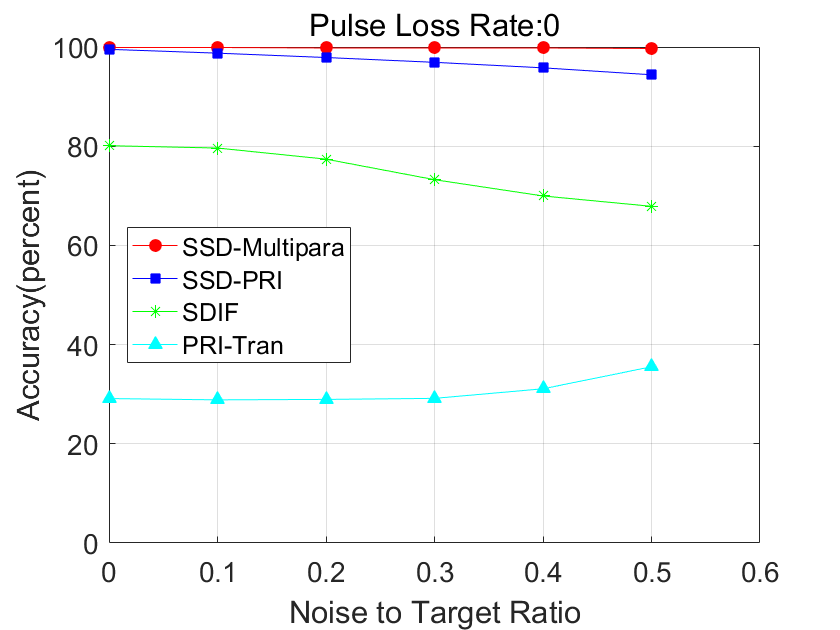}%
\label{exp4b}}
\hfil
\subfloat[]{\includegraphics[width=2.3in]{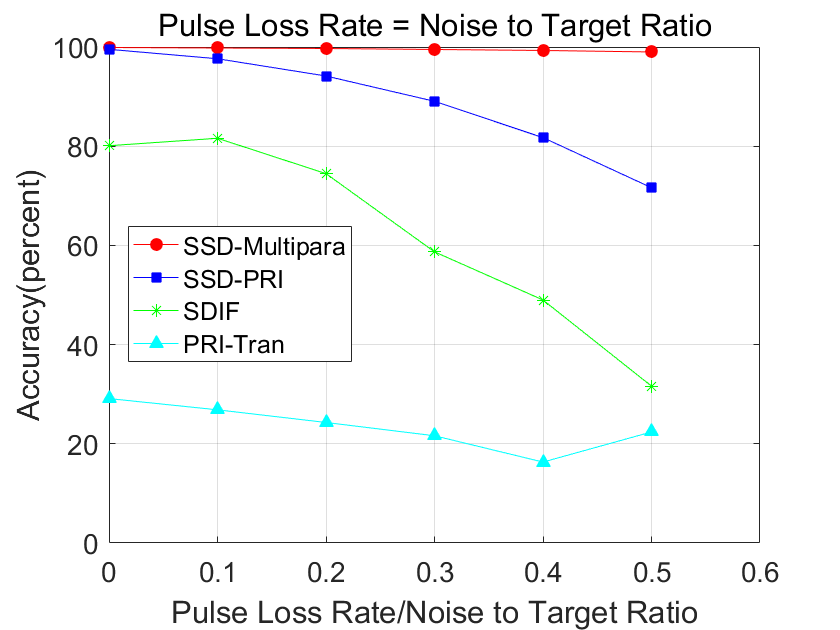}%
\label{exp4c}}
\caption{Test results of Experiment 4.}
\label{exp4_res}
\end{figure*}

In addition, we tested the trained model on datasets generated under three different conditions: 1) there is pulse loss, but no random noise pulse, i.e., ${\rho _n} = 0$; 2) there is no pulse loss, i.e., ${\rho _l} = 0$, but there are random noise pulses; 3) there are both pulse loss and random noise pulses, and ${\rho _l} = {\rho _n}$. The results are shown in Figs. \ref{exp1_res}–\ref{exp4_res}.

The experimental results confirmed the feasibility and efficiency of the SSD-Multipara method.

1) Experiments 1 and 2 show that PW and PRI can be used for radar signal deinterleaving simultaneously.

2) Experiment 1 shows that, although the SSD-PRI method cannot deinterleave pulses based on PRI in this case, the SSD-Multipara method can deinterleave pulses based on features generated by the coupling of PW and PRI, and obtain excellent deinterleaving results.

3) Experiment 2 shows that, although the PW values of the three targets overlap, the features generated by the coupling of PW and PRI significantly improve the deinterleaving efficiency.

4) Experiment 3 shows that RF and PRI can be used for radar signal deinterleaving simultaneously. The three RF variation modes: constant RF, agile RF among pulses, and agile RF among pulse groups, when coupled with PRI, can significantly improve the deinterleaving efficiency.

5) Experiment 4 shows that PA and PRI can be used for radar signal deinterleaving simultaneously. The three PA variation modes: Radar non-scanning, Radar mechanical-scanning, and Radar phase-scanning, when coupled with PRI, can significantly improve the deinterleaving efficiency.

6) The results obtained from the control experiments indicate that the SDIF and PRI-TRAN methods often do not achieve the best deinterleaving accuracy, even when the data quality is the best. This is because the threshold was adjusted to achieve the best overall performance of these methods on the obtained samples, which also reflects the shortcomings of such methods. In addition, in this type of method, the pulses remaining after deinterleaving are regarded as noise pulses. Therefore, in some cases, when the proportion of noise pulses increases, the deinterleaving accuracy increases. However, in fact, the deinterleaving accuracy of target pulses decreases.

\section{CONLUSION}
Using the concept of semantic segmentation and the features of multiple parameters obtained from radar signals, a new radar signal deinterleaving method employing NNs was investigated in this paper. The following conclusions can be drawn:

1) Compared to the PRI-based method, the SSD-Multipara method exhibits significantly better deinterleaving efficiency.

2) Compared to the SSD-PRI method, the SSD-Multipara method adapt to the situation where the PRI features of multiple radars are the same.

3) Compared to existing radar signal multiparameter-based deinterleaving methods, the SSD-Multipara method can adapt to complex radar signal features and signal environments, and can use different parameters in one step.

4) In this paper, the SSD method was extended from utilizing one parameter to two parameters. Presumably, the method can be further extended to include more parameters, i.e., the input data dimension can be $3*N$  or $4*N$ . In this way, the radar signals can be divided into more classes, and improved deinterleaving results can be achieved.

5) When multiple targets in the pulse stream do not have distinguishable features in the PRI, PW, RF, PA, and other parameters, the radar signal deinterleaving method based on semantic segmentation will encounter a bottleneck. Under such conditions, can we directly identify the target in the pulse stream without deinterleaving the pulse stream? This will be our future research direction.

\bibliography{Library}

\ifCLASSOPTIONcaptionsoff
  \newpage
\fi

\end{document}